\documentclass[12pt]{article}
\pdfoutput=1

\usepackage{subfigure}
\usepackage{amsmath,amssymb,amsthm,amsfonts,array,bbm}
\usepackage{graphicx}
\usepackage{color}
\usepackage{hyperref}
\usepackage{cite}
\usepackage[american]{babel}
\usepackage{slashed}
\usepackage{caption}

\newcommand{\ti}{\widetilde}
\newcommand{\tr}{\textrm{Tr} \,}

\newcommand{\wat}{\widehat}
\newcommand{\ol}{\overline}

\newcommand{\N}{\mathcal{N}}
\newcommand{\p}{\partial}
\newcommand{\be}{\begin{equation}} \newcommand{\ee}{\end{equation}}
\newcommand{\bea}{\begin{equation} \begin{aligned}} \newcommand{\eea}{\end{aligned} \end{equation}}

\newcommand{\sD}{\slashed{D}}

\newcommand{\dd}{\text{d}}

\newcommand{\cH}{\mathcal{H}}
\newcommand{\cI}{\mathcal{I}}

\newcommand{\cM}{\mathcal{M}}
\newcommand{\cN}{\mathcal{N}}

\newcommand{\cQ}{\mathcal{Q}}

\newcommand{\bC}{\mathbb{C}}

\newcommand{\bN}{\mathbb{N}}

\newcommand{\bR}{\mathbb{R}}
\newcommand{\bZ}{\mathbb{Z}}

\numberwithin{equation}{section}

\begin{document}

\begin{titlepage}

\vspace*{-2cm} 
\begin{flushright}
	{\tt  CERN-TH-2018-257} 
\end{flushright}
	
	\begin{center}
		
		\vskip .5in 
		\noindent

		{\Large \bf{Note on Monopole Operators in Chern-Simons-Matter Theories}}
		
		\bigskip\medskip
		
		Benjamin Assel\\
		
		\bigskip\medskip
		{\small 
			Theory Department, CERN, CH-1211, Geneva 23, Switzerland
		}
		
		\vskip .5cm 
		{\small \tt benjamin.assel@gmail.com}
		\vskip .9cm 
		{\bf Abstract }
		\vskip .1in
\end{center}
	
\noindent
Monopole operators in Chern-Simons theories with charged matter have been studied using the state-operator map in CFTs, as states on $\bR\times S^2$ with background magnetic flux on $S^2$. Gauge invariance requires a dressing with matter modes which provides non-zero spin to the monopoles. In this note we propose a description of the monopole operators directly on $\bR^3$, as a singular behavior of the gauge and matter fields in the vicinity of the insertion point, with a dressing. We study abelian theories with a charged boson or a charged fermion. We extend the discussion to abelian supersymmetric Chern-Simons-matter theories and describe the BPS monopoles, which have spin and preserve a single supercharge. We match our results against the prediction from the superconformal index. 

\noindent

\vfill

\end{titlepage}

\setcounter{page}{1}

\tableofcontents

\section{Introduction}
\label{sec:intro}

Monopole operators in three-dimensional Euclidean gauge theories are local operators whose insertion is defined by prescribing a Dirac monopole singularity at the insertion point for a $U(1)$ gauge field, with $U(1)$ embedded into the gauge group \cite{Borokhov:2002ib,Borokhov:2002cg,Borokhov:2003yu}. In an abelian theory with gauge field $A$, for an insertion at the origin in $\bR^3$, the monopole configuration is
\be
F = \dd A = -\frac{q}{2} \star \dd \Big( \frac 1r \Big) \,, \quad  q \in \bZ \,,
\label{MonopSing}
\ee
with $r$ the radial coordinate in $\bR^3$. Dirac quantization of the magnetic flux imposes $q\in \bZ$. The monopole singularity is spherically symmetric and carries no spin. As a local operator it has charge $q$ under the global $U(1)$ topological symmetry whose conserved current is $j = \star F$.  Monopole operators are rather ubiquitous in studies of three-dimensional gauge theories. For instance, they can serve as order parameter for symmetry broken phases in second order phase transitions \cite{Read:1989zz,Read:1990zza,Senthil1490,2004PhRvB..70n4407S}, they are dual to elementary fields under dualities \cite{Aharony:2015mjs,Seiberg:2016gmd,Benini:2017dus,Komargodski:2017keh,Benini:2017aed,Jensen:2017bjo,Benini:2018umh}, they are crucial to understanding the vacuum structure of supersymmetric theories \cite{Cremonesi:2013lqa}, they arise in non-perturbative corrections to superpotentials \cite{Affleck:1982as,Aharony:1997bx}, they can be used to deform the action and trigger RG-flows leading to infrared dualities \cite{Aharony:1997gp,Benvenuti:2016wet,Benini:2017dud,Benvenuti:2017kud}, they participate in infrared symmetry enhancement mechanisms \cite{Benini:2018bhk}. The higher dimensional cousin of the monopole operator is the most studied 't Hooft loop in four dimensional gauge theories.  Although this is not necessary, one can ``dress" a monopole insertion with uncharged matter fields inserted at the same point. This plays an important role in monopole operator counting problems in non-abelian theories \cite{Cremonesi:2013lqa}.

In a theory with Chern-Simons term at level $k\in \bZ$, a monopole of magnetic charge $q$ acquires a gauge charge $kq$. To obtain a gauge invariant operator one must dress the monopole with the insertion of {\it charged} matter fields. This is usually understood by using the state-operator map in a CFT. Under this map, monopole operators are described as states of the theory on the cylinder $\bR\times S^2$, in the sector where there is a magnetic flux of charge $q$ on $S^2$. The dressing arises as excitations of the matter fields in this monopole background \cite{Chester:2017vdh}. Because of the monopole background, the spacetime spin of the charged fields is modified. This phenomenon manifests itself in the form of the ``monopole spherical harmonics" studied in \cite{Wu:1976ge,Wu:1977qk}, which are the spherical harmonics of a scalar field of charge one in the $q$-monopole background. The smallest angular momentum of the scalar field harmonics is $\frac{|q|}{2}$.
In this context, the excitations of a scalar field dressing a monopole background provide non-zero spin to the gauge invariant monopole states, and the corresponding monopole operators of the CFT on $\bR^3$ have non-trivial spin.

It is not obvious how to realize these monopole operators as local operators on $\bR^3$. In particular we cannot simply insert charged fields $\phi(x)$ at the insertion point, since on one hand it would not account for the extra spin carried by the dressing, and on the other hand regular solutions to the equation of motions have the charged fields set to zero at the location of the monopole insertion. No such issue arises for dressing with uncharged matter fields.

In this note we overcome these difficulties and propose a realization of the monopole operator insertions by giving a singular behavior to the charged matter fields and dressing the monopole with matter modes that appear in the singular expansion.  We consider abelian gauge theories with minimally coupled matter fields. Our primary example is the theory of a single charged boson, which we study in Section \ref{sec:CSbtheory}. Our prescription is to require a profile for the scalar field $\phi$ that satisfies its equation of motion in the vicinity of the insertion point and to impose Gauss's law, which arises from the gauge field equation of motion. We find that these constraints can be met only if one allows for profiles where $\phi$ and $\ol\phi$ are related by a modified conjugation relation, with $\phi$ diverging at the origin and $\ol\phi$ vanishing at the origin, or vice-versa. The relation between $\phi$ and $\ol\phi$ can be understood as inherited from the standard complex conjugation of the Lorentzian theory on $\bR\times S^2$. The profiles for $kq>0$ take the form
\bea
\phi & \ = \  \sum_{j,m} a_{jm} r^{-\beta_j - \frac 12} Y_{qjm}  \cr 
\ol\phi & \ = \  \sum_{j,m}  \ol a_{jm} r^{\beta_j-\frac 12}  \ol Y_{qjm}  
\label{SingProfileIntro}
\eea
with $\beta_j =  \frac 12 \sqrt{(2j+1)^2 - q^2}$ and $Y_{qjm}$ are the monopole harmonics with background magnetic charge $q$. The allowed values of the angular momentum are $j = \frac{|q|}{2} + n$, $n \in \bZ_{\ge 0}$. The moduli $|a_{jm}|$ are constrained by Gauss's law, while the phases $e^{i\lambda_{jm}}$, defined by $a_{jm}=|a_{jm}| e^{i\lambda_{jm}}$, are dynamical variables.

Removing a ball of radius $\epsilon$ around the insertion point and requiring a boundary term on the $S^2$ boundary that cancels the boundary piece of the bulk field variation, we find that in the $\epsilon \to 0$ limit the monopole insertion must be dressed with a factor $\prod_{j,m} (e^{-i\lambda_{jm}})^{n_{jm}}$, where $n_{jm}$ is a collection of positive integers satisfying $\sum_{jm} n_{jm} =kq$ and in terms of which the $|a_{jm}|$ moduli are fixed (see Equations \ref{GenSingProfile},\ref{GenGL},\ref{GenDressing}). This dressing term also restores the gauge invariance of the monopole insertion in the Chern-Simons theory.

Because the dressing factors $e^{-i\lambda_{jm}}$ have spin $j$, the monopoles transform in non-trivial $Spin(3)$ representations, which are easily worked out in this simple theory. Our findings mimic to a large extent the construction of monopole states of the theory on $\bR\times S^2$ in Hamiltonian quantization as presented in \cite{Chester:2017vdh}.

We extend our construction to the abelian theory with a charged fermion field in Section \ref{sec:CSftheory}. The only qualitative difference with respect to the scalar field case is that the ``occupation numbers" $n_{jm}$ of each fermion mode take values $0$ or $1$ only, due to the fermionic statistic.

\medskip

We then carry on to study supersymmetric monopoles in Section \ref{sec:Susy}. The simplest instance arises in supersymmetric $\cN=2$ SQED theory with a single charged chiral multiplet and Chern-Simons level $k$. To define a supersymmetric monopole insertion one starts by requiring a half-BPS monopole singularity \eqref{BPSmonop}. In this background the singular profiles for the boson and fermion fields are slightly modified (or rather simplified). By studying the BPS conditions we find that $\frac{1}{4}$-BPS monopoles can be constructed if one restricts the singular profiles and dressing factors to certain scalar and fermion modes which obey BPS conditions (see Equation \eqref{GenProfileSusyFinal}).  A schematic summary of the construction of BPS monopoles is given in Table \ref{tab:BPSmonop}. In an infrared CFT, the BPS monopoles belong to short multiplets at threshold $A_1$, or $\ol A_1$, in the language of \cite{Cordova:2016emh}.
We compute the quantum numbers of the BPS monopoles, including their dimension, as a function of the infrared R-charges of the fields. We compare our results with the superconformal index of the abelian SQED theory and find an exact agreement. 

The analysis in this paper should be understood as performed in a small coupling limit of the theories, which is the large $k$ limit (or the small $g_{YM}$ coupling limit if one uses a Yang-Mills UV regulating term). However most of the qualitative results, such as the monopole operator content and their spin representations, does not change as continuous couplings are turned on, including quadratic and quartic scalar potentials, which might be fine-tuned to reach infrared interacting CFTs.\footnote{The Chern-Simons level is not a continuous coupling, but it is likely that the spectrum of monopoles that we describe is correct at any value of $k$. This is certainly the case for BPS monopoles.} Because they can acquire large anomalous dimensions, it is difficult to assess the fate of non-supersymmetric monopole operators along RG-flows in strongly coupled theories (small $k$).

Although we study only simple abelian theories, the construction presented in this note should generalize to other abelian theories and non-abelian theories without major modifications.

\bigskip

\noindent{\bf Acknowlegments}: \quad 

I thank Stefano Cremonesi for valuable comments on the draft and Mark Mezei for discussions related to the topic.

\section{CS theory with a charged boson}
\label{sec:CSbtheory}

We consider a $U(1)$ Chern-Simons theory at level $k\in \bZ$ with a complex scalar field $\phi$ of charge one. The action is
\be
S = \frac{ik}{4\pi} \int A\wedge \dd A + \int\dd x^3  D^\mu\ol\phi D_\mu \phi \,,
\ee
with $D_\mu\phi = (\p_\mu -iA_\mu)\phi$. Often one introduces a Yang-Mills term which acts as a UV regulator to the action. In this discussion we will not need it.\footnote{Also we do not introduce potential terms in the action (mass term and quartic potential). The idea is that we study the local operators in the limit of vanishing couplings, to make things simple. Our findings however will not depend on continuous deformations of the action and will be valid, for instance, in the critical theory with a quartic interaction, which flows to a non-trivial IR fixed point.}
The equations of motion (eom) are
\bea
& (i) \quad   \frac{k}{2\pi} F + \star(\ol\phi D\phi - \phi D\ol\phi) =0 \,, \cr
& (ii) \quad  D^2\phi =0 \,.
\label{CSbEOM}
\eea
To define a local monopole operator of magnetic charge $q$, we require that the gauge field has a Dirac monopole singularity at the origin in Euclidean space
\be
\frac{1}{2\pi} \int_{S^2} F = q \in \bZ \,,
\ee
which, in a convenient gauge, corresponds to the gauge field profile
\bea
A &= \frac q 2 (\pm 1 - \cos \theta)\dd\varphi \,, \cr
F &= \dd A = \frac{q}{2}\sin\theta \, \dd\theta\dd\varphi =  -\frac{q}{2} \star \dd \Big( \frac 1r \Big) \,, \quad  q \in \bZ \,,
\label{MonopSingbis}
\eea
where we used the spherical coordinates $r\ge 0$, $\theta \in [0,\pi]$, $\varphi \sim \varphi + 2\pi$.
We will denote $\omega_2 = \sin\theta \, \dd\theta\dd\varphi$ the volume form of the unit $S^2$.

The eom $(i)$ implies Gauss's law constraint on the magnetic flux emanating from the origin
\be
 \int_{S^2} \star(\ol\phi D\phi - \phi D\ol\phi)  = -\frac{k}{2\pi} \int_{S^2} F = - kq  \,.
 \label{GaussLaw}
\ee
Therefore we need to require a certain profile at the origin for the field $\phi$ as well, compatible with Gauss's law. We will require a profile compatible with the eom $(ii)$ close to the origin.

Equation $(ii)$ is modified (compared to the free scalar theory) by the presence of the non-trivial gauge connection.
To solve for the eom $(ii)$ we must use the so-called monopole scalar harmonics $Y_{qjm}$ on $S^2$ \cite{Wu:1976ge}, which are not functions but sections of the gauge bundle over $S^2$. They are eigenfunctions of the modified Laplacian on the sphere with magnetic background flux  $q$,
\bea
& \vec L_q^2 Y_{qjm} = j(j+1) Y_{qjm} \,, \quad L_{q,z} Y_{qjm} = m Y_{qjm} \,, \cr
&L_{q,z} := -i\p_\varphi - q \,, \quad  \vec L_q^2 := - \nabla_{S^2}^2 + \frac{2q}{\sin^2\theta} (\cos\theta -1)L_z \,,
\eea
with $m=-j,-j+1,\cdots,j-1,j$. They obey the orthonormal relations $\int_{S^2} \omega_2 \ol Y_{qjm}Y_{q'j'm'} = \delta_{qq'}\delta_{jj'}\delta_{mm'}$ and $\ol Y_{qjm} := Y_{qjm}^\ast = (-1)^{q+m} Y_{-q,j,-m}$.
An important difference with respect to the standard scalar harmonics is that the allowed values for $j$ are $j=\frac{|q|}{2} + n$, for $n\in \bZ_{\ge 0}$. The smallest value is $j =\frac{|q|}{2}$. 

Using the ansatz $\phi = g(r) Y_{qjm}(\theta,\varphi)$ and writing the Laplacian in spherical coordinates, $\nabla^2\phi  = \frac{1}{r^2}\p_r(r^2\p_r\phi) + \frac{1}{r^2}\nabla^2_{S^2}$, the eom $(ii)$ reduces to
\be
0 = -\p_r(r^2 \p_r g(r)) - \frac{q^2}{4} g(r) + j(j+1)g(r) \,.
\ee
This admits a singular and a regular solution $g(r) = r^{-\beta_j - \frac 12}$ and $g(r) = r^{\beta_j-\frac 12}$ respectively, with
\be
\beta_j = \frac 12 \sqrt{(2j+1)^2 - q^2}  \ > \frac 12 \,.
\label{beta}
\ee
We can thus solve the eom $(ii)$ with the scalar profiles
\be
\phi = \frac{a}{r^{\beta_j + \frac 12}} Y_{qjm} \,,
\ee
or 
\be
\phi = a  r^{\beta_j-\frac 12} Y_{qjm}  \,,
\ee
with $a\in \bC$.
However, if we adopt the standard reality condition $\ol \phi = \phi^\ast$, we find that no such profile can solve Gauss's law \eqref{GaussLaw}. Actually, with $\ol\phi = \phi^\ast$ the left-hand-side of \eqref{GaussLaw} is imaginary, whereas we need it real.
Instead, if we think of $\phi$ and $\ol\phi$ as independent fields of charge $1$ and $-1$ respectively, with the equations of motions $D^2\phi = 0$, $D^2\ol\phi = 0$, we have the solutions 
\be
\phi = \frac{a}{r^{\beta_j+\frac 12}} Y_{qjm} \,, \quad \ol\phi = \bar a r^{\beta_j-\frac 12} \ol Y_{qjm} \,,
\ee
with
\be
 \int_{S^2} \star(\ol\phi D\phi - \phi D\ol\phi) = -2\beta_j a \ol a \,.
\ee
Thus Gauss's law is satisfied with
\be
a \ol a = \frac{kq}{2\beta_j} \,.
\ee
While we do satisfy Gauss's law, we remark that the solutions do not satisfy the local equations $(i)$ in \eqref{CSbEOM}. Trying to impose a solution to $(i)$ seems a too strong requirement and we find that imposing only Gauss's law, which is the integrated equation, will fit our purposes.\footnote{See also section 3.1.4 in \cite{Chester:2017vdh} for a comment and a possible explanation on this point.} 
We conclude that, assuming $kq>0$, one should impose the profiles at the origin
\bea
\phi &= \frac{e^{i\lambda}}{r^{\beta_j+\frac 12}}\sqrt{\frac{kq}{2\beta_j}} Y_{qjm} + \text{sub} \,, \cr
\ol\phi &= e^{-i\lambda}r^{\beta_j-\frac 12} \sqrt{\frac{kq}{2\beta_j}} \ol Y_{qjm} + \text{sub} \,,
\label{SingProfile}
\eea
where the phase $e^{i\lambda}$, $\lambda \sim \lambda + 2\pi$, is a fluctuating field and ``sub" denotes subleading terms in small $r$. The angle $\lambda$ cannot be chosen as a fixed background because it transforms under gauge transformations:
\be
A \to A + \dd\Lambda \,, \quad \lambda \to \lambda + \Lambda(0) \,,
\label{lambdatransfo}
\ee
where $\Lambda(0)$ is the evaluation at the origin of the gauge parameter.
If $kq <0$, one has to exchange the roles of $\phi$ and $\ol\phi$, i.e. changing $\beta_j \to -\beta_j$ in the profiles \eqref{SingProfile}.

The surprising feature of the profiles \eqref{SingProfile} is that the leading behavior of $\ol\phi$ is related to that of $\phi$ by an unusual reality condition $\ol\phi = r^{2\beta_j}\phi^\ast$. In general one can think of $\ol\phi$ and $\phi$ as independent complex fields and define a half-dimensional slice in field space to integrate on in the path integral. The choice of slice should make the action positive definite. The standard choice is $\ol\phi =\phi^\ast$, but there could be other choices. 
We will not study how to choose a slice, or define proper reality conditions, compatible with \eqref{SingProfile} and simply assume that it can be done. We notice that if we perform a conformal map to the cylinder $\bR\times S^2$ and a Wick rotation to Lorentzian signature, the reality condition that we observe in \eqref{SingProfile} becomes usual complex conjugation.
We will also not study the precise form of the subleading term ``sub".
\medskip

When introducing a diverging profile at the origin, a common procedure is to cut a small ball $B_{\epsilon}$ of radius $\epsilon>0$ and allow for a boundary term on $S^2_\epsilon = \p B_\epsilon$. Such a boundary term is fixed, in principle, by requiring a well-defined variation principle, namely the cancellation of boundary terms coming from the field variation of the action. 
The variation of the scalar action produces the boundary term
\bea
\delta S |_{\rm bdy} &= -\int_{S^2_\epsilon} \omega_2 \epsilon^2 (\delta\ol\phi \,  \p_r\phi + \p_r\ol\phi \, \delta \phi) \cr
&= - \int_{S^2_\epsilon} \omega_2 i kq |Y_{qjm}|^2 \delta\lambda + O(\epsilon^\alpha) \cr
&=  -  i kq \delta\lambda + O(\epsilon^\alpha)  \,,
\eea
with $\alpha>0$. To cancel this term (in the limit $\epsilon \to 0$) we should thus add the boundary term\footnote{In our conventions the integrand of the path integral is $e^{-S - S_{\rm bdy}}$.}
\be
S_{\rm bdy} = i kq\int_{S^2_\epsilon} \omega_2  |Y_{qjm}|^2 \lambda  = ikq\lambda   \,.
\label{SbdyWithMonop}
\ee
Therefore we find that we should insert $e^{-ikq\lambda}$ at the origin to complete the operator insertion.\footnote{Describing this insertion as a local operator insertion in terms of $\phi$ and $\ol\phi$ is not convenient.}
This result agrees nicely with the analysis of the gauge invariance of the monopole operator. Let us explain this point.

Under a field variation $A \to A +\delta A$ the Chern-Simons action changes by a bulk plus a boundary term
\be
\delta S_{\rm CS} = \frac{ik}{2\pi} \int \delta A\wedge F + \frac{ik}{4\pi}\int_{S^2_{\epsilon}} \delta A \wedge A \,.
\ee
In order for the boundary term $\delta A \wedge A$ to cancel we can fix one component of the gauge field to zero, say $A_{\theta}=0$, on the boundary \cite{Moore:1989yh}. This is compatible with the presence of the Dirac monopole singularity. The variation $\delta S_{\rm CS}$ reduces only to the bulk term
\be
\delta S_{\rm CS} = \frac{ik}{2\pi} \int \delta A\wedge F  \,.
\ee
Specializing to a gauge transformation $\delta A = d\Lambda$, we find
\be
\delta_\Lambda  S_{\rm CS} = \frac{ik}{2\pi} \int \dd\Lambda \wedge F = -\frac{ik}{2\pi} \int_{S^2_\epsilon} \Lambda F  \,.
\ee
We see that $\delta_\Lambda  S_{\rm CS}$ is a boundary term which is non-vanishing in the presence of a non-zero magnetic flux \eqref{MonopSingbis}. In the limit $\epsilon \to 0$, we have
\be
\delta_\Lambda  S_{\rm CS} = - i k q \Lambda(0) \,.
\label{CSgaugetransfo}
\ee
Note that the transformation is compatible with $\Lambda$ being $2\pi$-periodic, since $kq \in \bZ$.  
This gauge variation is nicely canceled by the gauge variation of the dressing factor $e^{-ikq\lambda}$,
\be
\delta_\Lambda (e^{-ikq\lambda} e^{-S_{\rm CS}}) = 0 \,.
\ee
The insertion of $e^{-ikq\lambda} = (e^{-i\lambda})^{kq}$ can be thought of as a dressing with $kq$ modes of $\ol\phi$ at the origin.

To summarize, with $kq >0$, a monopole insertion at the origin $M_q(0)$ is defined in the path integral formulation by requiring the Dirac monopole singularity \eqref{MonopSingbis}, the scalar profiles \eqref{SingProfile} and the insertion of $e^{-ikq\lambda}$, with $\lambda$ the phase defined in \eqref{SingProfile}. For $kq <0$, the scalar profiles are exchanged ($\beta_j \to -\beta_j$) and the dressing $e^{-ikq\lambda} =(e^{i\lambda})^{-kq}$ can be thought of as a dressing with $-kq$ modes of $\phi$ at the origin.
\bigskip

\noindent{\bf Generalization and spin of monopoles}
\medskip

As such this operator does not transform nicely under $Spin(3)$ rotations. If we label $\lambda = \lambda_{jm}$ the phase appearing in the profile \eqref{SingProfile}, we observe that $e^{i\lambda_{jm}}$ (or $e^{-i\lambda_{jm}}$) transforms as a component of the spin $j$ representation.  The dressing operators $e^{-ikq\lambda_{jm}} = (e^{-i\lambda_{jm}})^{kq}$ however do not transform in a representation of the rotation group, or rather transform into operators that we have not yet discussed . To find the missing operators, we need to generalize the monopole insertions.

The generalization goes as follows. Assuming $kq >0$, we require, in addition to the monopole flux singularity \eqref{MonopSingbis}, the scalar field profile at the origin
\bea
\phi &= \sum_{j\ge \frac{|q|}{2}}\sum_{m=-j}^j  \frac{a_{jm}}{r^{\beta_j+\frac 12}} Y_{qjm} + {\rm sub} \,, \cr
\ol\phi &= \sum_{j\ge \frac{|q|}{2}}\sum_{m=-j}^j  \ol a_{jm} r^{\beta_j-\frac 12} \ol Y_{qjm}  + {\rm sub} \,,
\label{GeneralSingProfile}
\eea
with $\ol a_{jm} = (a_{jm})^\ast$, and with only a finite number of non-zero $a_{jm} \in \bC$. 

Gauss's law \eqref{GaussLaw} imposes the constraint
\be
\sum_{j\ge \frac{|q|}{2}}\sum_{m=-j}^j 2\beta_j |a_{jm}|^2 = kq \,.
\label{GenGLconstr}
\ee
The boundary contribution in the variation of the action is now canceled by adding the boundary term
\bea
S_{\rm bdy} &=  \sum_{j\ge \frac{|q|}{2}}\sum_{m=-j}^j  \int_{S^2_\epsilon} \omega_2 |Y_{qjm}|^2 i (2\beta_j)  |a_{jm}|^2  \lambda_{jm}  \cr
 & =
\label{GenSbdy}   \sum_{j\ge \frac{|q|}{2}}\sum_{m=-j}^j i (2\beta_j) |a_{jm}|^2  \lambda_{jm} \,.
\eea
Now recall that the phases $\lambda_{jm}$ are $2\pi$ periodic, so, for the boundary term $\exp(-S_{\rm bdy})$ to make sense, we must impose the quantization conditions  $2\beta_j |a_{jm}|^2 := n_{jm} \in \bN$ for all $j,m$. 
To satisfy Gauss's law \eqref{GenGLconstr} one must then choose a collection of non-negative integers $n_{jm}$, such that $\sum_{j,m} n_{jm} = kq$. The scalar profiles become
\bea
\phi &= \sum_{j\ge \frac{|q|}{2}}\sum_{m=-j}^j  \frac{e^{i\lambda_{jm}}}{r^{\beta_j+\frac 12}} \sqrt{\frac{n_{jm}}{2\beta_j}} Y_{qjm} + {\rm sub} \,, \cr
\ol\phi &= \sum_{j\ge \frac{|q|}{2}}\sum_{m=-j}^j   e^{-i\lambda_{jm}} r^{\beta_j-\frac 12} \sqrt{\frac{n_{jm}}{2\beta_j}} \ol Y_{qjm}  + {\rm sub} \,,
\label{GenSingProfile}
\eea
with 
\be
\underline{\text{Gauss's law}}: \qquad  \sum_{j\ge \frac{|q|}{2}}\sum_{m=-j}^j n_{jm} = kq \,,  \quad (n_{jm} \in \bZ_{\ge 0} ) 
\label{GenGL}
\ee
and the monopole insertion is completed by the dressing operator
\be
\underline{\text{Dressing term}}: \qquad  \prod_{j\ge \frac{|q|}{2}} \prod_{m=-j}^j \exp(-i n_{jm}\lambda_{jm}) \,.
\label{GenDressing}
\ee
Once again the dressing operator restores the gauge invariance of the monopole insertion, since the phases $\lambda_{jm}$ all transform as $\lambda_{jm} \to \lambda_{jm} + \Lambda(0)$ under a gauge transformation.

This defines the insertion at the origin of a monopole operator $M_{q {\bf n}}$ with ${\bf n} = (n_{jm})$ with $n_{jm} \ge 0$ and $|{\bf n}| := \sum_{jm} n_{jm} = kq$. As is sometimes done in the literature, one can think heuristically of the dressing as the insertion  of $kq$ factors, where each factor is thought of as a $\p^n\ol\phi$ insertion at the origin, with $\ol\phi$ an operator of spin $\frac{|q|}{2}$ and $n = j-\frac{|q|}{2}$. This is however more of a book-keeping device rather than a correct statement.

For $kq<0$ we need to invert the roles of $\phi$ and $\ol\phi$ by exchanging the profiles in \eqref{GenSingProfile} (this is implemented by $\beta_j \to -\beta_j$, $n_{jm} \to - n_{jm}$) and dressing the insertion with $-kq$ factors $e^{i\lambda_{jm}}$. Heuristically we dress the monopole with modes of $\phi$ instead of $\ol\phi$. Gauss's law becomes in this case 
\be
\sum_{j\ge \frac{|q|}{2}}\sum_{m=-j}^j n_{jm} = -kq > 0  \,.
\ee
\medskip

We can now reconsider the question of the transformation under $SU(2)= Spin(3)$ rotations. From the definition of the profiles \eqref{GenSingProfile} we understand that the operators $e^{-i\lambda_{jm}}$ for $|m| \le j$ form a spin $j$ representation, which we denote ${\bf j}$. 

We deduce that the set of operators $M_{q{\bf n}}$ with fixed $n_j := \sum_{m=-j}^j n_{jm}$ (satisfying $\sum_j n_j = kq$), transform in the tensor product representation $\bigotimes\limits_{j\ge \frac{|q|}{2}} [\mathbf{j}^{\otimes n_{j}}]_{\rm sym}$, where $[...]_{\rm sym}$ takes the symmetric product of the factors in the bracket. 
\be
\left\lbrace M_{q{\bf n}} \, \Big| \,  \sum_{m=-j}^j n_{jm} = n_j \right\rbrace  \quad \longrightarrow \quad  SU(2)  \ \text{rep} \quad \bigotimes\limits_{j\ge \frac{|q|}{2}} [\mathbf{j}^{\otimes n_{j}}]_{\rm sym}    \,.
\ee
This is a reducible representation.\footnote{It contain the symmetric traceless representations and the traces.} In the minimal case $k=q=1$, the occupation numbers ${\bf n}$ of the monopoles $M_{q{\bf n}}$ have a single non-zero entry $n_{jm}=1$. In this case the $2j+1$ monopoles with $n_j=1$ form a spin $j$ representation, the smallest spin being $j=\frac q2 = \frac 12$.

\bigskip

These monopole operators, with these spin quantum numbers, exist in the theory deformed by a scalar mass term and scalar quartic potential, since continuous deformations do not affect the $SU(2)$ representations in which the monopole transform. We can think about the monopole operators in the theory with critical quadratic and quartic interactions, which flows to an infrared CFT.  Ideally one would like to know the conformal dimension of these monopoles in the CFT. 
Computing the dimension of the monopole operators (or any unprotected operator) is a notoriously hard problem in a strongly coupled field theory. In the limit of large number of charged fields $N_f \gg 1$ the infrared theory is effectively weakly coupled and the monopole dimension can be computed pertubatively in $1/N_f$ \cite{Borokhov:2002ib,Pufu:2013vpa} (see also a $d=4-\epsilon$ approach in \cite{Chester:2015wao}). In the Chern-Simons theory, 't Hooft-like limits were considered (with large $k,N_f$ or large $k,N_c$ and fixed ratio). The monopole dimension is then extracted from the leading contribution to the free energy of the theory on $S^1_\beta\times S^2$ in the small temperature limit $\beta \to \infty$ \cite{Radicevic:2015yla,Chester:2017vdh,Dyer:2013fja}, sometimes relying on numerical evaluations. These results are not directly applicable to the theory of a $U(1)$ gauge group with a single flavor.\footnote{ In the large $k$ limit, the saddle point analysis of \cite{Chester:2017vdh} should be valid, even with a single charged scalar, which corresponds to taking large $\kappa$. The results presented for the bosonic theory are numerical. We were not able to isolate a result which applies to our situation.}

\section{CS theory with a charged fermion}
\label{sec:CSftheory}

We consider a $U(1)$ gauge theory with Chern-Simons kinetic term at level $k\in \bZ$ and a fermion $\psi$ of $U(1)$ charge 1. The Euclidean action is
\be
S = \frac{ik}{4\pi} \int A\wedge \dd A + \int d^3x \,  i \ol\psi \sD \psi \,,
\ee
with $D_\mu\psi = (\nabla_\mu - i A_\mu)\psi$. The fermions $\psi,\ol\psi$ have two complex components $\psi_{\alpha}, \ol \psi_{\alpha}$, $\alpha=1,2$. 
In the path integral formulation of the theory, we adopt a regularization of the fermion determinant in the background of the gauge field $A$ that produces a phase $\exp(-\frac{i\pi}{2}\eta(A))$, where $\eta(A)$ is the APS eta invariant \cite{Atiyah:1975jf,Atiyah:1976jg,Atiyah:1980jh}. 
We refer to \cite{Witten:2015aba} for an in-depth discussion. If we add a mass term with parameter $m$ and integrate out the fermion field, the low-energy effective Chern-Simons level is $k_{IR} = k - \frac 12 + \text{sgn}(m)\frac 12$ \cite{Seiberg:2016gmd}. This corresponds to a one-loop quantum correction to the bare Chern-Simons level and should be understood as the physical Chern-Simons level. For a massless fermion we have a bare Chern-Simons term at level $k$ and the phase given by the APS eta invariant. This is often imprecisely referred to as the theory at level $k-\frac 12$. 

The equation of motions (eom) are
\bea
& (i) \quad  \frac{ik}{2\pi} (\star F)_\mu + \ol\psi \gamma_u\psi  = 0 \,, \cr
& (ii) \quad  \sD \psi =0 \,.
\label{CSfEOM}
\eea
We define the Dirac monopole singularity with magnetic charge $q \in \bZ$ as in \eqref{MonopSingbis}. In the fermionic theory Gauss's law  stems from eom $(i)$ and is given by
\footnote{If we had considered a massive fermion, the Chern-Simons level appearing in Gauss's law should be replaced by the one-loop corrected level $k_{IR} \in \bZ$ discussed above. For a massless fermion the situation is more subtle. Consistency with the monopole spectrum obtained in radial quantization indicates a posteriori that the bare Chern-Simons level $k$ is the one appearing in Gauss's law (and not the half-integer $k-\frac 12$ for instance). We do not fully understand this point.}
\be
kq  = \frac{k}{2\pi} \int_{S^2} F =  i \int_{S^2} \star(\ol\psi \gamma_\mu\psi \, \dd x^\mu ) = i \int_{S^2} \omega_2 r^2 \, \ol\psi \gamma_r \psi  \,.
\label{GLfermion}
\ee
Therefore we should supplement the Dirac monopole singularity with a singular fermion profile such that this constraint is satisfied.
To accomplish this, we study the solutions of the eom $(ii)$ is the vicinity of the monopole insertion. The solutions are expressed in terms of the so-called spin $\frac 12$ monopole harmonics on $S^2$, which were studied in \cite{Borokhov:2002ib, Dyer:2013fja}. In the notation of \cite{Dyer:2013fja} (Appendix A), there are two types of spin $\frac 12$ harmonics, explicitly given by\footnote{We correct here some typos in \cite{Dyer:2013fja}} \footnote{This form assumes a frame where $\gamma^a$, $a=1,2,3$, are the standard Pauli matrices. }
\bea
& T_{qjm}(\hat n) = \left(
\begin{array}{c}
\sqrt{\frac{j+m}{2j}} Y_{q,j-\frac 12,m-\frac 12}(\hat n) \cr
\sqrt{\frac{j-m}{2j}} Y_{q,j-\frac 12,m+\frac 12}(\hat n) 
\end{array} \right)
\,, \quad \text{for} \quad j \ge \frac{|q|}{2} + \frac 12 \,, \cr
& S_{qjm}(\hat n) =  \left(
\begin{array}{c}
-\sqrt{\frac{j-m+1}{2j+2}} Y_{q,j+\frac 12,m-\frac 12}(\hat n) \cr
\sqrt{\frac{j+m+1}{2j+2}} Y_{q,j+\frac 12,m+\frac 12}(\hat n)
\end{array} \right)
 \,, \quad \text{for} \quad j \ge \frac{|q|}{2} - \frac 12 \,,
\eea
where $\hat n$ is a unit vector parametrizing $S^2$ (replacing $(\theta,\varphi)$), and $|m| \le j$. In the notation of \cite{Borokhov:2002ib} they correspond to $\phi_{j-\frac 12,j,m} = T_{qjm}$, $\phi_{j+\frac 12,j,m} = S_{qjm}$, obeying $\vec L^2 \phi_{\ell j m} = \ell(\ell+1)\phi_{\ell j m}$, $\vec J^2 \phi_{\ell j m} = j(j+1)\phi_{\ell j m}$, $J_3 \phi_{\ell j m} = m\phi_{\ell j m}$. 
Importantly the $T_{qjm}$ harmonics exist only for  $j \ge \frac{|q|}{2} + \frac 12$, and the $S_{qjm}$ harmonics only for $j \ge \frac{|q|}{2} -\frac 12$, with the understanding that $j -  \frac{|q|}{2} + \frac 12 \in \bZ$.

A solution to the eom with total angular momentum $(j,m)$ is of the form 
\be
\psi = t(r) T_{qjm} + s(r) S_{qjm} \,, 
\ee
for $ j \ge \frac{|q|}{2} + \frac 12$, or simply $\psi = s(r) S_{qjm}$ for $j = \frac{|q|}{2} -\frac 12$. Solving the eom in cylindrical coordinates around the origin is a little tedious. Fortunately this was accomplished in \cite{Borokhov:2002ib}. The final result is\footnote{The equations are solved for the theory on the cylinder in \cite{Borokhov:2002ib}. The solutions in flat space are easily obtained by performing the Weyl rescaling back from $\bR\times S^2$ to $\bR^3$, taking into account that $\psi$ has scaling dimension one, $\psi_{\bR^3} = \frac 1r \psi_{\bR\times S^2}$.}
\be
\underline{j=|\frac q2| -\frac 12}: \quad \psi = \frac{a}{r} S_{qjm} \,, 
\ee
with $a\in \bC$ and $|m| \le j$ (this corresponds to zero-modes of the theory on $\bR\times S^2$), and
\bea
\underline{j \ge \frac{|q|}{2} +\frac 12}:  \quad  \psi  &=  a_1 r^{\beta_j-1} \Big[ -\frac{q}{2} T_{qjm} + ( j + \frac 12 - \beta_j) S_{qjm} \Big] \cr
& \ \ +  a_2 r^{-\beta_j-1} \Big[ \frac{q}{2} T_{qjm} + (j + \frac 12 + \beta_j) S_{qjm} \Big]  \,,
\eea
with $|m| \le j$, $\beta_j$ given in \eqref{beta}, and $a_1,a_2 \in \bC$ two constants. 
The solution with $a_2=0$ goes to zero at $r=0$, while the solution with $a_1=0$ diverges at $r=0$.

As in the bosonic case, the standard conjugation relation $\ol\psi=\psi^\dagger$ does not lead to solutions of Gauss's law \eqref{GLfermion}.
For $j=|\frac q2| -\frac 12$, we can solve \eqref{GLfermion} close to the origin by requiring a profile
\be
\quad \psi =\frac{a}{r} S_{q,|\frac q2| -\frac 12,m} + \text{sub} \,, \quad  \ol\psi = - i \psi^\dagger  \,.
\ee
For $j \ge \frac{|q|}{2} +\frac 12$, it is not possible to solve \eqref{GLfermion} by assuming a profile solving the eom and having $\ol\psi \propto \psi^\dagger$. Instead we can consider the profiles
\bea
& \psi = a r^{-\beta_j-1} \Big[ \frac{q}{2} T_{qjm} + ( j + \frac 12 + \beta_j ) S_{qjm} \Big] +  \text{sub} \,, \cr
& \ol\psi = i \ol a r^{\beta_j-1} \Big[ \frac{q}{2} \ol T_{qjm} - ( j + \frac 12 - \beta_j )  \ol S_{qjm} \Big] +  \text{sub} \,,
\eea
with $\ol T_{qjm} = T_{qjm}^\dagger$ and $\ol S_{qjm} = S_{qjm}^\dagger$. 

The modulus $|a|= (a\ol a)^{1/2}$ is fixed by solving Gauss's law as a function of $k$, $q$ and $j$. 
Using the computations of the appendix of \cite{Borokhov:2002ib},\footnote{To reproduce these computations, one would need at some point to compute integrals of three $Y$ functions which are given in terms of Wigner $3j$ symbols, as found in \cite{Wu:1977qk}.} we find that the fermion profile is
\bea
& \psi =  e^{i\lambda} r^{-\beta_j-1} \sqrt{\frac{k}{j+\frac 12}} \Big[ \frac{q}{2} T_{qjm} + (j + \frac 12 + \beta_j ) S_{qjm} \Big] + \text{sub} \,, \cr
& \ol\psi = i e^{-i\lambda} r^{\beta_j-1} \sqrt{\frac{k}{j+\frac 12}} \Big[ \frac{q}{2} \ol T_{qjm} - ( j + \frac 12 - \beta_j) \ol S_{qjm} \Big] + \text{sub} \,,
\label{FermionProfile}
\eea
where it is understood that $T_{qjm}=0$ for $j= \frac{|q|}{2} -\frac 12$.

Because  the profile of $\psi$ is divergent we can think of regularizing the operator by cutting a ball $B_{\epsilon}$ of radius $\epsilon>0$ around the origin and study the limit $\epsilon \to 0$. Varying the fermion action at finite $\epsilon$ produces a boundary term
\bea
\delta S_{\rm fermion} |_{\rm bdy} &= \int_{S^2_\epsilon}  \omega_2 \epsilon^2 \ol \psi\gamma^r \psi \, \delta\lambda \cr
&\xrightarrow{\epsilon \to 0} -i kq \delta\lambda \,,
\eea
where $S^2_\epsilon = \p B_{\epsilon}$ and we have used the fact that the fermion background satisfies Gauss's law \eqref{GLfermion} to reach the final result in the limit $\epsilon \to 0$. To cancel this boundary term (in the limit $\epsilon \to 0$) we add to the operator insertion the dressing term
\be
e^{-i kq \lambda} \,.
\ee
As in the scalar theory, the dressing factor has gauge charge $-kq$, compensating for the gauge transformation of the Chern-Simons term and restoring the full gauge invariance of the monopole operator insertion.

There is however an issue with the dressing factor. The phase $e^{-i\lambda}$, as defined by the fermion profile \eqref{FermionProfile}, is a Grassmann-odd field. Therefore it vanishes when raised to a power two or bigger. Thus the dressing term $(e^{-i\lambda})^{kq}$ vanishes, except for $kq=1$. To be able to define monopole operators with higher values of $kq$ we need more fermion modes.
\bigskip

\noindent{\bf Generalization}
\medskip

The monopole insertion can be generalized. We assume $kq >0$. Since this is analogous to the scalar field case, we only go through the main lines, skipping details. We can require a singular profile of the fermion field
\bea
& \psi = \sum_{j \ge \frac{|q|}{2} -\frac 12}\sum_{m=-j}^j  e^{i\lambda_{jm}} r^{-\beta_j-1} \sqrt{\frac{n_{jm}}{q(j+\frac 12)}}\Big[ \frac{q}{2} T_{qjm} + ( j + \frac 12 + \beta_j) S_{qjm} \Big] + \text{sub} \,, \cr
& \ol\psi = \sum_{j \ge \frac{|q|}{2} -\frac 12}\sum_{m=-j}^j i  e^{-i\lambda_{jm}} r^{\beta_j-1} \sqrt{\frac{n_{jm}}{q(j+\frac 12)}} \Big[ \frac{q}{2} \ol T_{qjm} - ( j + \frac 12 - \beta_j) \ol S_{qjm} \Big] + \text{sub} \,,
\label{GenFermionProfile}
\eea
where it is understood that $T_{qjm} = 0$ if $j=|\frac q2| -\frac 12$, and $n_{jm}$ are positive and satisfy $\sum_{j \ge \frac{|q|}{2} -\frac 12}\sum_{m=-j}^j  n_{jm} = kq$.
The required dressing term is then
\be
\underline{\text{Dressing term}}: \quad   \prod_{j\ge \frac{|q|}{2} -\frac 12} \prod_{m=-j}^j \exp(-i n_{jm}\lambda_{jm}) \,,
\ee 
The periodicity of the phases $\lambda_{jm} = \lambda_{jm} + 2\pi$ implies $n_{jm} \in \bZ_{\ge 0}$.
Because $\exp(-i\lambda_{jm})$ are Grassmann-odd fields, there should at most one power of each such factor in the dressing operator. This means that $n_{jm} \in \{0,1\}$. 
\medskip

This defines the insertion of the monopole operator $\ti M_{q {\bf n}}$ with ${\bf n} = (n_{jm})$, satisfying
\be
\underline{\text{Gauss's law}}: \quad   \sum_{j \ge \frac{|q|}{2} -\frac 12}\sum_{m=-j}^j n_{jm} = kq  \,, \quad  n_{jm} \in \{0,1\}  \,.
\ee

\noindent{\bf Spin of the monopoles}
\smallskip

From the definition of the fermion profile we observe that the phases $e^{i\lambda_{jm}}$ transform in the spin $j$ representation. 
It follows that the set of monopole operators $\ti M_{q {\bf n}}$ with fixed $n_j := \sum_{m=-j}^j n_{jm}$ transform in the $SU(2)$ representation $\bigotimes_{j \ge  \frac{|q|}{2} -\frac 12} [{\bf j}^{\otimes n_j}]_{\rm anti-sym}$, where $[...]_{\rm anti-sym}$ takes the anti-symmetric product of the factors in the bracket. 
In the minimal case where $k=q=1$, the bare monopole has gauge charge one and a single occupation number $n_{jm}$ is non-zero for each dressed monopole. The $2j+1$ monopoles $\ti M_{q {\bf n}}$ with $n_j=1$ transform in the spin $j$ representation, and the minimal spin is $j=\frac q2 - \frac 12 = 0$ corresponding to a scalar operator.

\medskip

For $kq<0$, we must exchange the roles of $\psi$ and $\ol\psi$, by exchanging the profiles in \eqref{GenFermionProfile} and dress the monopole singularity with $-kq$ factors of $e^{i\lambda_{jm}}$. Heuristically we dress the monopole with modes of $\psi$, instead of $\ol\psi$. Gauss's law becomes in this case
\be
\sum_{j \ge \frac{|q|}{2} -\frac 12}\sum_{m=-j}^j n_{jm} = -kq  \,, \quad  n_{jm} \in \{0,1\} \,.
\ee

\section{BPS monopoles in $\cN=2$ Chern-Simons SQED}
\label{sec:Susy}

Supersymmetric Chern-Simons theories with charged matter admit many monopole operators which can defined by allowing bosonic and fermionic singular profiles.
With the minimal amount of supersymmetry, 3d $\cN=1$ gauge theories do not admit monopole operators preserving supersymmetries. This is simply because the Dirac monopole background \eqref{MonopSingbis} breaks all supercharges.
With $\cN=2$ supersymmetry, it is still true that the gauge field monopole background breaks all supersymmetries, but it is possible to preserve half of them by requiring a singular behavior for the scalar field in the vector multiplet.
An $\N=2$ vector multiplet  is composed of the fields $(A,\sigma,\lambda,\ol\lambda,D)$: a gauge field $A$, a real scalar $\sigma$, a two-components fermion $\lambda$ and a real auxiliary scalar $D$. We provide the supersymmetry transformations of the abelian vector multiplet in Appendix \ref{app:susytransfo}.
The BPS equations are
\bea
 0 &= \delta\lambda = \frac{i}{2}\epsilon^{\mu\nu\rho} \gamma_\mu \varepsilon  F_{\nu\rho} - D\varepsilon + i \gamma^\mu\varepsilon \p_\mu \sigma \cr 
&= i \gamma^\mu\varepsilon ((\star F) + \dd \sigma)_\mu -D\varepsilon  \,, \cr
 0 &= \delta\ol\lambda = \frac{i}{2}\epsilon^{\mu\nu\rho} \gamma_\rho \ol\varepsilon  F_{\mu\nu} + D\varepsilon - i \gamma^\mu\ol\varepsilon \p_\mu \sigma \cr
 &=  i \gamma^\mu\ol\varepsilon ((\star F) - \dd \sigma)_\mu + D\ol\varepsilon  \,, \cr
\eea
where $\varepsilon$ and $\ol\varepsilon$ are two independent two-component complex spinors, parametrizing the $\cN=2$ supersymmetry transformations with generators $Q_\alpha$ and $\ol Q_{\alpha}$, $\alpha=1,2$.
 
 A half-BPS monopole operator of magnetic charge $q\in\bZ$ is defined by imposing the Dirac monopole singularity for the gauge field and a singular profile for $\sigma$ \cite{Borokhov:2002cg}:
\bea
\underline{\text{BPS monopole}}: \quad   F &= -\frac{q}{2} \star \dd \Big( \frac 1r \Big) + O(1) \,,  \cr
\sigma &= u\frac{q}{2r} + O(1) \,,
\label{BPSmonop}
\eea
with $u\in \{+1,-1\}$ and $D=0$.  This profile obeys $\star F + u \dd\sigma = 0$, preserving the supercharges $Q_\alpha$, $\alpha=1,2$ for $u=+1$ and $\ol Q_{\alpha}$, $\alpha=1,2$, for $u=-1$. 

\medskip

In the absence of a Chern-Simons term  the monopole profile \eqref{BPSmonop} defines a gauge invariant local half-BPS chiral operator.\footnote{The generalization to a non-abelian Yang-Mills theory does not present difficulties and is well-known \cite{Borokhov:2003yu}.} 
We now consider an abelian theory with Chern-Simons term at level $k$. The $\cN=2$ Chern-Simons action is
\be
S_{CS}^{\cN=2} = \frac{ik}{4\pi} \int d^3x  \Big[\epsilon^{\mu\nu\rho} A_\mu \p_\nu A_\rho - \ol\lambda \lambda + 2 D\sigma \Big]\,.
\ee
In the absence of Yang-Mills (or rather Maxwell) kinetic term, $\sigma$ and $\lambda$ are auxiliary fields.
We consider the theory with a single chiral multiplet with matter fields $(\phi,\psi,F)$ of $U(1)$ charge $+1$, comprising a complex scalar $\phi$, a two-component complex spinor $\psi$ and a complex auxiliary scalar  $F$, with action
\bea
S_{\rm chi} = \int d^3x \Big[ & D^\mu\ol\phi D_\mu\phi - i \ol\psi\gamma^\mu D_\mu\psi + \ol F F \cr
& +\ol\phi(\sigma^2 + iD)\phi + i \ol\psi\sigma\psi + i\ol\psi\lambda\phi - i \ol\phi\ol\lambda\psi  \Big]\,.
\eea
To define the insertion of a monopole operator of magnetic charge $q$ we can proceed as described in the previous sections by requiring a Dirac monopole singularity for the gauge field and a singular profile for the charged matter fields $\phi$ and/or $\psi$. This leads to a variety of monopole operators with various spins that we do not analyze here. All these monopole break all supersymmetries as explained.
\medskip

To find a supersymmetric monopole in the $\cN=2$ Chern-Simons theory, we start by requiring a half-BPS singular profile \eqref{BPSmonop} with magnetic charge $q\in \bZ$, for the vector multiplet fields. 
We take $u=-1$ and look for monopole operators that preserve $\ol Q_\alpha$.

To define the monopole operators we must require a singular profile for matter fields. 
The analysis of the scalar field profiles is only slightly modified compared to section \ref{sec:CSbtheory}. The modification arises because of the coupling of $\phi$ to $\sigma$. In the vicinity of the origin the equation of motion for $\phi$ is now
\be
D^\mu D_\mu \phi - \frac{q^2}{4r^2}\phi = 0 \,,
\ee
due to the singularity \eqref{BPSmonop} of $\sigma$. The  effect of the extra term is to simplify the parameter $\beta_j$ to $j+\frac 12$, making the radial profiles of the solutions that of a free scalar field. The rest of the analysis is unaffected.\footnote{When we regularize the insertion by removing a ball $B_{\epsilon}$, we must add extra boundary terms to preserve supersymmetry. A standard calculation shows that we must add $S^{\rm susy}_{\rm bdy} = \int_{S^2_\epsilon}\omega_2 \epsilon^2 ( i\ol\psi\gamma_r\psi - \ol\phi D_r \phi)$. This term evaluates to a constant in the limit $\epsilon\to 0$ and its variation  $\delta S^{\rm susy}_{\rm bdy}$ is subleading in $\epsilon$ compared to the other boundary terms entering in our discussion, therefore we can neglect it.}
\medskip

A similar modification occurs for the fermion modes, which are analysed as in Section \ref{sec:CSftheory}. The equation of motion for $\psi$ in the supersymmetric background becomes 
\be
i\sD \psi + i\frac{q}{2r}\psi = 0 \,.
\ee
The solutions are then of the form $\psi = a_1 r^{j-\frac 12} T_{qjm} + a_2 r^{-j-\frac 32} S_{qjm}$. This can be worked out from a small modification of the analysis in \cite{Borokhov:2002ib}.\footnote{Mapping these solutions to the theory on $\bR\times S^2$, we reproduce the spectrum of fermion mode energies of $\cN=2$ SQED computed in \cite{Borokhov:2002cg}.}
\medskip

We thus find that the supersymmetrization of the monopole singularity simplifies the expressions, as one might have expected.
To solve Gauss's law, we must choose the boson backgrounds 
\bea
& \phi_{jm} = r^{-j-1} Y_{qjm} \,, \cr
& \ol\phi_{jm} = r^j \ol Y_{qjm} \,,
\label{SusyBosonModes}
\eea
or permutations of $\phi$ and $\ol\phi$, with $j \ge \frac{|q|}{2}$, and the fermion backgrounds
\bea
& \psi_{jm} = r^{j-1/2} T_{qjm} \,, \cr
& \ol\psi_{jm} = r^{-j-3/2} \ol S_{qjm} \,,
\label{SusyFermionModes}
\eea
or permutations of $\psi$ and $\ol\psi$, with $j \ge \frac{|q|}{2} + \frac 12$.
For $j=|\frac q2|- \frac 12$ there is only the $S_{qjm}$ mode for $\ol\psi$ which is not paired with a $T_{qjm}$ mode for $\psi$, therefore we do not allow this mode.

\medskip

We are interested in monopole operators preserving supercharges $\ol Q_\alpha$, so we must select the bosonic and fermionic profiles solving the $\ol Q$-BPS equations in the background of a BPS monopole singularity \eqref{BPSmonop}.
Let us look at the scalar field. The BPS equations are
\bea
& 0 =\delta_{\ol\varepsilon} \psi= \ol\varepsilon F \,, \cr
& 0 = \delta_{\ol\varepsilon}\ol\psi =  i \gamma^\mu\ol\varepsilon D_\mu\ol\phi + i \ol\varepsilon \sigma\ol\phi \,.
\label{BPSeqn1}
\eea
The first equation imposes $F=0$ at the origin.\footnote{We must only require that the BPS equations hold as we approach the origin.}
To evaluate the right-hand-side of the second equation we need the explicit form of the monopole scalar harmonics $Y_{qjm}$, given in \cite{Wu:1976ge}.
With $x:= \cos\theta$, we have\footnote{$Y_{jm}$ are not functions but sections of the gauge line bundle on $S^2$. This is the value on a patch covering one hemisphere of $S^2$: the patch where $A = \frac 12(1-\cos\theta)\dd\varphi$. On the other patch the value is multiplied by $e^{-iq\varphi}$.}
\be
Y_{qjm}(x,\varphi) = C_{jm}  e^{i(m + \frac q2)\varphi} (1-x)^{\frac q4 + \frac m2} (1+x)^{-\frac q4 + \frac m2} \p_x^{j+m} [(1-x)^{-\frac q2 + j} (1+x)^{\frac q2 + j} ] \,,
\ee
with $C_{jm}$ real constants. We take the gamma matrices to be $\gamma^1=\tau^3$, $\gamma^2=\tau^1$, $\gamma^3=\tau^2$, with $\tau^i$ the standard Pauli matrices, and we look for solutions preserving the supercharge $\ol Q_2$ generated by $\ol\varepsilon=\binom{0}{1}$. We find the two BPS equations
\bea
(a) \quad  & 0 = \p_r\ol\phi + \frac{(1-x)(1+x)}{rx}\p_x\ol\phi + \frac{q}{2rx}\ol\phi \,, \cr
(b) \quad & 0 = \p_r\ol\phi - \frac{x}{r}\p_x\ol\phi - \frac{i}{r(1-x)(1+x)}\big( \p_\varphi \ol\phi + i\frac{q}{2}(1-x)\ol\phi \big) \,.
\eea
We now evaluate these BPS equations on the $(j,m)$-component of the $\ol\phi$ profile, $\ol\phi_{jm} =  r^{j} \ol Y_{qjm}$. We find
\bea
(a) \quad   0 &= \ol Y_{qjm} (j - m) + \ol Y_{qj,m+1}\frac{C_{qjm}}{C_{qj,m+1}} x^{-1} (1-x)^{\frac 12}(1+x)^{\frac 12} e^{i\varphi}   \,, \cr
(b) \quad  0 &= \ol Y_{qjm} ( j - m )   - \ol Y_{qj,m+1}\frac{C_{qjm}}{C_{qj,m+1}} x (1-x)^{-\frac 12}(1+x)^{-\frac 12} e^{i\varphi} \,.
\eea
These equations are solved for $m=j$, in which case $\ol Y_{qj,m+1}=0$. 
We conclude that the profile with $m=j$ preserves the supercharge $\ol Q_2$. The other values $m \neq j$ do not solve the BPS equation for $\ol Q_2$.

Alternatively we may look for solutions preserving the supercharge $\ol Q_1$ generated with $\ol\epsilon = \binom{1}{0}$. 
The analysis goes along the same lines and we find that among the profiles $\ol\phi_{jm} = r^j \ol Y_{qjm}$, only the choice $m=-j$ solves the BPS equations for $\ol Q_1$. There is no background $\ol\phi_{jm}$ preserving both $\ol Q_\alpha$ supercharges.
\medskip

Let us turn to fermion field. The BPS equations are
\bea
& 0 = \delta_{\ol\varepsilon}\phi = \ol\varepsilon\psi  \,, \cr
& 0 = \delta_{\ol\varepsilon} \ol F = \ol\varepsilon (i\gamma^\mu D_\mu\ol\psi - i \ol\psi \sigma) \,.
\eea
The equations on the second line are solved for all $\ol\psi$ profiles since they are the equations of motion for $\ol\psi$ and the profiles are defined as solutions to the equations of motion. The equation on the first line is very simple. It admits solutions for some $\ol\epsilon$ only if the two components in $\psi$ are proportional. The background $\psi_{jm}$ in \eqref{SusyFermionModes} is explicitly
\be
\psi_{jm} \propto T_{qjm}(\hat n) = \left(
\begin{array}{c}
\sqrt{\frac{j+m}{2j}} Y_{q,j-\frac 12,m-\frac 12}(\hat n)\cr
\sqrt{\frac{j-m}{2j}} Y_{q,j-\frac 12,m+\frac 12}(\hat n)
\end{array} \right) 
\,.
\ee
We observe that the two components of $\psi_{jm}$ are independent except when one is vanishing. This happens for $m=j$, when the lower component vanishes, and for $m=-j$ when the upper component vanishes. Explicity the profiles $\psi_{jj}$ preserve $\ol Q_2$ and the profiles $\psi_{j,-j}$ preserve $\ol Q_1$. Here again there is no solution preserving both $\ol Q_\alpha$ supercharges. 
\medskip

If we had chosen the diverging backgrounds $r^{-j-\frac 32} S_{qjm}$ for $\psi$ (instead of the converging $r^{j-\frac 12} T_{qjm}$), we would have found no solution to the BPS equation. 
Similarly if we had exchanged the roles of $\phi_{jm}$ and $ \ol\phi_{jm}$ in \eqref{SusyBosonModes} we would have found no $\ol Q$-BPS solutions.

\bigskip

Finally we should ask whether the dressing factors preserve supersymmetries or not. 
The dressing operator is a product of terms $e^{-i \lambda_{jm}}$ and $e^{i \wat\lambda_{jm}}$ which arise as the coefficients of the modes $\ol\phi_{jm}$ and $\psi_{jm}$ respectively. The supercharges act on $e^{-i \lambda_{jm}}$ and $e^{i \wat\lambda_{jm}}$ as they act on the modes $\ol\phi_{jm}$ and $\psi_{jm}$. Since $\delta_{\ol\varepsilon}\ol\phi =0$ and $\delta_{\ol\varepsilon}\psi = \ol\varepsilon F =0$ (at the origin), the dressing factors are $\ol Q_\alpha$-invariant for both $\alpha=1,2$.

From the BPS computations above one can also conclude that the profiles $\phi_{jj}$ and $\psi_{jj}$ (and thus the modes $e^{-i \lambda_{jj}}$ and $e^{i \wat\lambda_{jj}}$) are not $\ol Q_2$-exact, and that similarly the profiles $\phi_{j,-j}$ and $\psi_{j,-j}$ (and the modes $e^{-i \lambda_{j,-j}}$ and $e^{i \wat\lambda_{j,-j}}$) are not $\ol Q_1$-exact. 

\bigskip

We conclude that the matter profiles can preserve at most one supercharge and, in order to do so, one should allow only for boson and fermion fields with $m = j$, or only those with $m= -j$. For instance the profiles preserving $\ol Q_2$ are
\bea
\underline{\ol Q_2\text{-BPS profiles}}: \quad \phi &= \sum_{j\ge \frac{|q|}{2}}  \frac{e^{i\lambda_{jj}}}{r^{j+1}} \sqrt{\frac{n_{jj}}{2j+1}} Y_{qjj} + {\rm sub} \,, \cr
\ol\phi &= \sum_{j\ge \frac{|q|}{2}}  e^{-i\lambda_{jj}} r^{j} \sqrt{\frac{n_{jj}}{2j+1}} \ol Y_{qjj}  + {\rm sub} \,, \cr
\psi &= \sum_{j\ge \frac{|q|}{2}+\frac 12} e^{\wat i\lambda_{jj}} r^{j-\frac 12} \sqrt{\frac{\wat n_{jj}(j+\frac 12)}{\beta_j}} T_{qjj} + {\rm sub} \,, \cr
\ol\psi &= \sum_{j\ge \frac{|q|}{2}+\frac 12}  i\frac{e^{-i\wat\lambda_{jj}}}{r^{j+ \frac 32}} \sqrt{\frac{\wat n_{jj}(j+\frac 12)}{\beta_j}} \ol S_{qjj}  + {\rm sub} \,.
\label{GenProfileSusyFinal}
\eea
The integers $n_{jj} \in \bZ$ and $\wat n_{jj} \in \{0,1\}$ satisfy Gauss's law
\be
\sum_{j \ge \frac{|q|}{2}} n_{jj} - \sum_{j \ge \frac{|q|}{2}+\frac 12} \wat n_{jj} = k'_{(-)} q \,,
\label{GLsusy1}
\ee
with 
\be
k'_{(u)} = k -\frac 12 - \text{sgn}(q)\frac u2 \,, \quad u \in \{-1,1\} \,.
\ee
$k'_u$ is the ``infrared" Chern-Simons level. The shift of the bare Chern-Simons level $k$ arises from the regularization of the fermion determinants. For a fermion of mass $m$, the IR level is $k -\frac 12 + \text{sgn}(m)\frac 12$, as discussed in section \ref{sec:CSftheory}. Here the role of the mass $m$ is played by the background $-\sigma = -u\frac{q}{2r}$, and sgn$(\sigma) = $ sgn$(q) u$.

Notice that the (mod 2) integers $\wat n_{jj}$ are weighted with a minus sign in Gauss's law. This is because we imposed a diverging profile for $\ol\psi$ (instead of $\psi$) and thus we are dressing with modes of $\psi$ instead of $\ol\psi$. Gauss's law imposes that the gauge charge of the dressing compensates for the gauge charge of the bare monopole, and the modes of $\psi$ have opposite gauge charge compared to the modes of $\ol\psi$.

The dressing factor is
\be
\prod_{j\ge \frac{|q|}{2}} e^{-i n_{jj}\lambda_{jj}} \prod_{j\ge \frac{|q|}{2}+\frac 12} e^{i \wat n_{jj}\wat\lambda_{jj}} \,.
\ee
This describes the insertion of supersymmetric monopole operators $\ol\cM_{q{\bf n \wat n}}$, which are $\ol Q_2$-BPS operators. The same construction with the selection of the $m=-j$ modes, leads to the definition of $\ol Q_1$-BPS operators. 
\medskip

In this calculation we have chosen $u=-1$ for the supersymmetric monopole singularity \eqref{BPSmonop} and found BPS monopoles preserving one $\ol Q$ supercharge. We can also pick the other choice $u=1$, in which case the BPS monopole singularity \eqref{BPSmonop} preserves the two supercharges $Q_\alpha$. The appropriate bosonic and fermionic backgrounds in that case are obtained by exchanging the roles of $\phi$ and $\ol\phi$ in \eqref{SusyBosonModes}, and of $\psi$ and $\ol\psi$ in \eqref{SusyFermionModes}, namely the diverging backgrounds are those of $\ol\phi$ and $\psi$. From the BPS equations we find that allowing only the modes with $m=j$ defines monopole operators preserving $Q_1$, while allowing the modes $m=-j$ defines monopole operators preserving $Q_2$. Gauss's law imposes the constraint on the bosonic numbers $n_{j,\pm j}$ and fermionic numbers $\wat n_{j, \pm j}$,
\be
- \sum_{j \ge \frac{|q|}{2}} n_{j,\pm j} + \sum_{j \ge \frac{|q|}{2}+\frac 12} \wat n_{j,\pm j} = k'_{(+)} q \,,
\label{GLsusy2}
\ee
where the $\pm$ sign is $+$ for monopoles preserving $Q_1$ and  $-$ for those preserving $Q_2$.

These features are qualitatively summarized in Table \ref{tab:BPSmonop}.

\begin{table}
\begin{tabular}{c|c|c|c}
BPS monop. & behavior at $r=0$  & BPS modes & Gauss's law \cr
 \hline 
$\begin{array}{c}
\star F = \dd\sigma  \cr
(u=-1)
\end{array}$ 
 & $\phi,\ol\psi$ div.; $\ol\phi,\psi \to 0$  & 
$\begin{array}{c}
 \ol Q_1: m=-j  \cr
\ol Q_2: m=j 
\end{array}$ 
& $\sum_j n_{jm} - \wat n_{jm} = k'_{(-)}q$ \cr
\hline
$\begin{array}{c}
\star F = -\dd\sigma  \cr
(u=1)
\end{array}$ 
 & $\phi,\ol\psi \to 0$; $\ol\phi,\psi$ div. & 
$\begin{array}{c}
 Q_1: m=j  \cr
Q_2: m=-j 
\end{array}$
& $\sum_j n_{jm} - \wat n_{jm} = -k'_{(+)}q$
\end{tabular}
\caption{Main features of BPS monopoles. Each monopole preserves a single supercharge.}
\label{tab:BPSmonop}
\end{table}

\medskip

Notice that the supersymmetric monopoles of the Chern-Simons theory always preserve a single supercharge and are therefore $\frac 14$-BPS local operators. This is different from supersymmetric monopoles of Yang-Mills theories (with zero infrared Chern-Simons term) which preserve two supercharges and are  half-BPS chiral operators.

\subsection{Quantum numbers and superconformal multiplets}
\label{ssec:QuantumNumbers}

Having understood which monopole operators are BPS, we would like to describe their quantum numbers and to explain to which short superconformal multiplet they belong, in the SCFT that is believed to exist in the infrared limit.

We focus on the ``bosonic " monopoles $\ol\cM_{q\bf n}$, defined with the $u=-1$ BPS monopole background and a dressing by bosonic modes (i.e. modes of $\ol\phi$) only, with occupation numbers ${\bf n}=(n_{jm})$. The fermion modes are all set to zero for these monopoles, $\wat n= 0$. These monopoles exist only for $k'_{(-)} q >0$, since Gauss's law is in this case $\sum_{j,m} n_{jm} = k'_{(-)} q$.  We will use here $k':=k'_{(-)}$ and assume $k'>0$ and $q>0$ for simplicity.
Generically the monopoles $\ol\cM_{q\bf n}$ do {\it not} preserve any supersymmetry, but some of them preserve $\ol Q_1$ or $\ol Q_2$.
\medskip

A simple class of monopoles preserving $\ol Q_2$ has $n_{jj} = k'q$ for a chosen $j$, and $n_{jm}= 0$ for all other $j,m$ pairs. Their counterpart monopoles preserving $\ol Q_1$ have $n_{j,-j} = k'q$ and $n_{jm}=0$ for all other $j,m$ pairs. 

They both belong to the same irreducible spin representation $[{\bf j}^{\otimes k'q}]'_{\rm sym}$ of $SU(2)$, where $[...]'_{\rm sym}$ denotes the symmetric traceless product of the representations. This is simply the spin $k'qj$ representation. Let us denote $J=k'qj$ and $V^{(j)}_{q m}$ , with $|m| \le J$, the monopoles in this representation. The $\ol Q_2$ invariant monopole is $V^{(j)}_{qJ}$ and the $\ol Q_1$ invariant monopole is $V^{(j)}_{q,-J}$.  The other monopoles $V^{(j)}_{qm}$ are built out of the monopoles $\cM_{q\bf n}$ with $n_{j} := \sum_m n_{jm} = k'q$ and $n_{j'}=0$ for $j '\neq j$.

The dimension of the monopoles $V^{(j)}_{qm}$ can be computed using the BPS properties of the BPS operators. It is a sum of two contributions
\be
\Delta(V^{(j)}_{qm}) = \Delta_{\rm bare} + \Delta_{\rm dressing} \,.
\ee
The contribution from the dressing factor is the sum of the dimension of each individual factor $e^{-i\lambda_{jm}}$. This is easily extracted from the definition of the $(j,m)$ mode: $\ol\phi \sim e^{-i\lambda_{jm}} r^{j} \ol Y_{qjm}$. In an $\cN=2$ SCFT, the dimension of the anti-chiral field $\ol\phi$ is related to its R-charge\footnote{Hopefully there will be no confusion between the R-charge $r$ and the radial coordinate $r$. Both notations are fairly standard.} $-r$: $\Delta(\ol\phi)= - R(\ol\phi) = r$. It follows that the dimension of the dressing mode is 
\be
\Delta(e^{-i\lambda_{jm}}) = j + r \,.
\ee
The dressing has $k'q$ modes of $\lambda_{jm}$, leading to 
\be
\Delta_{\rm dressing} = k'q \big(j + r \big) = J + k'qr\,.
\ee
The contribution $\Delta_{\rm bare}$ from the bare anti-chiral monopole is also related to its R-charge by the BPS condition and can be computed as a sum of zero point energy of all oscillators in the theory on the cylinder \cite{Borokhov:2002cg}. It is given by \cite{Imamura:2011su,Benini:2011cma}
\be
\Delta_{\rm bare} = \frac{1-r}{2}q \,,
\ee
and the total dimension is
\be
\Delta(V_{qm}) =  J  + \big( k' - \frac 12 \big)qr +\frac{q}{2} \,.
\ee

Similarly the $U(1)$ R-charge of the BPS monopoles are computed as the sum of the R-charge of the bare monopole $R_{\rm bare} = -\frac{1-r}{2}q$ and the R-charge of the dressing $R_{\rm dressing} = k'q R(\ol\phi) = -kqr$,
\be
R(V^{(j)}_{qm}) =  - \big( k' - \frac 12 \big) qr -\frac{q}{2} \,.
\ee
The exact values of the R-charge and dimension of such BPS operators at the infrared fixed point depend on the $U(1)$ R-charge at this fixed point. This may not coincide with the UV R-charge, but rather is a combination of $U(1)_{R,{\rm UV}}$ with the $U(1)$ global symmetries of the IR fixed point. The parameter $r$ refers to the charge of $\phi$ under this infrared R-symmetry. This can be determined by extremizing the $S^3$ partition function of the $\cN=2$ theory under consideration \cite{Jafferis:2010un}.
\medskip

We recover that the BPS monopoles obey the BPS condition \\
 $\Delta = J_3 - R$ for $V^{(j)}_{q, J}$ and $\Delta = -J_3 - R$ for $V^{(j)}_{q, -J}$. This follows from the fact that the bare monopole and the dressing factors all obey the corresponding BPS condition.
 Another way to find the BPS monopoles is to consider only those dressed with modes $\ol\phi_{jm},\psi_{jm}$ (or $\phi_{jm},\ol\psi_{jm}$) which obey such BPS conditions.
 
 The fact that these monopoles obey the BPS condition means that they define non-trivial elements of the $\ol Q_1$ or $\ol Q_2$ cohomologies and therefore contribute to the superconformal index defined with $\ol Q_1$ or with $\ol Q_2$.
 
\bigskip

The monopoles $V^{(j)}_{qm}$ are only the simplest BPS monopoles.
These considerations apply in general to the BPS monopoles dressed with both bosonic and fermionic modes, as described in the previous subsection.  We leave as a exercise to work out the quantum numbers in this general case.

\bigskip

\noindent{\bf Superconformal multiplet}
\medskip

In the infrared SCFT the $\ol Q_{\alpha}$-BPS monopoles belong to short superconformal mulitplets called $A_1$ in the classification of \cite{Cordova:2016emh} (or $\chi_S$ in \cite{Minwalla:2011ma}). These are the only short superconformal multiplets in 3d $\cN=2$ SCFTs which accommodate for non-zero spin. They are $\frac 14$ BPS operators in the sense that they are non-trivial in $\ol Q_1$ cohomology or in $\ol Q_2$ cohomology, but not in both. Moreover they are not the bottom component of the $A_1$ multiplet, but rather descendants. Indeed the bottom component $C$ in this multiplet satisfies $\Delta = J - R + 1$, where $J$ is the $SU(2)$ spin and $R$ is the $U(1)$ R-charge. In components we can write it $C_{\alpha_1\cdots\alpha_{2j}}$, for the operator with spin $j$. It satisfies the shortening condition $\ol Q_\beta C^{\beta}{}_{\alpha_1\cdots \alpha_{2j-1}} = 0$.\footnote{If $j=0$, $C$ is rather the bottom component of an $A_2$ multiplet with shortening condition $\ol Q_\alpha \ol Q^\alpha C=0$.} The descendants $D_{\alpha_1,\cdots, \alpha_{2j+1}} = \ol Q_{(\alpha_1} C_{\alpha_2\cdots\alpha_{2j+1})}$ obeys $\Delta = J - R$. The BPS monopoles are identified with the components $D_{11\cdots 1}$ and $D_{22\cdots 2}$, which are non-trivial in $\ol Q_2$ and $\ol Q_1$ cohomology respectively. They are the only operators in the full multiplet contributing to the superconformal index defined with $\ol Q_2$ or with $\ol Q_1$ (see \cite{Minwalla:2011ma} for details).

There is a mirror discussion for $Q_\alpha$-BPS monopoles defined with $u=+1$ backgrounds and $\phi,\ol\psi$ dressing modes.

Note in particular that this is different from supersymmetric monopoles in Maxwell (or Yang-Mills) theory without Chern-Simons term, which are not  dressed with charged matter field. There, the monopoles are chiral ($\frac 12$-BPS) operators with no spin. They are the bottom components of $B_1$ (or $\ol B_1$) superconformal multiplets \cite{Cordova:2016emh} and are non-trivial in both $\ol Q_\alpha$ (or both $Q_\alpha$) cohomologies.

\subsection{Superconformal index}
\label{ssec:SCI}

We can compare our findings with the superconformal index of the theory \cite{Bhattacharya:2008zy}, which counts BPS operators in the theory. For a given choice of supercharge $\cQ$ one can define an index as a trace over the Hilbert space $\cH$ of the theory on the cylinder $\bR \times S^2$, refined with fugacities associated to generators that commute with $\{\cQ,\cQ^{\dagger}\}$. For 3d $\cN=2$ theories there are two indices \cite{Minwalla:2011ma}, that can be defined with the choices $\cQ = Q_1$ or $\cQ = Q_2$ (choosing $\ol Q_\alpha$ supercharges yield the same indices). To adapt to common conventions, we consider the index selecting states which obey $H- J_3 - R =0$. This is achieved by choosing $\cQ = Q_1$, 
\be 
\cI = \tr_{\cH} (-1)^F e^{-\beta \{Q_1,Q_1^{\dagger} \}} x^{H +J_3} w^{Q_m} \,,
\ee
with $H$ the energy generator, $J_3$ the Cartan generator of the $Spin(3)=SU(2)$ rotations on $S^2$, $F$ the fermion number, and $Q_m$ the generator of the topological symmetry, which counts the magnetic flux on $S^2$. There is no other global symmetry in our SQED theory with a single chiral multiplet, because the flavor symmetry is gauged in this model.
Only states satisfying with $\{Q_1, Q_1^{\dagger} \}:= H - J_3 - R  = 0$ contribute to this index, so that it is independent of $\beta$.

Under the state-operator correspondence, the index $\cI$ counts BPS operators of the theory on $\bR^3$ which are non-trivial elements of the $Q_1$ cohomology. They satisfy $\Delta = J_3 + R$, where the dilatation operator $\Delta$ is identified with the Hamitonian $H$ of the cylinder theory.
\medskip

The superconformal index can be computed elegantly by supersymmetric localization as the partition function of the theory on $S^1\times S^2$ with periodic boundary conditions for the fermions around $S^1$ and some background deformations accounting for the parameters $x,w$ \cite{Imamura:2011su,Kapustin:2011jm}. For the SQED theory with a single chiral multiplet it takes the form
\be
\cI = \sum_{q \in \bZ} (-1)^{k'_+q} w^{q} \oint \frac{\dd z}{2\pi i z} z^{k'_+q} \, x^{(1-r)\frac{|q|}{2}} \prod_{n\ge 0} \frac{ 1- z^{-1} x^{|q|+2-r + 2n}}{1 - z x^{|q| + r + 2n}} \,,
\ee
where $k'_+ = k - \frac 12 - \text{sgn}(q)\frac 12$ is the infrared effective Chern-Simons level.

This is a sum over magnetic sectors weighted with $w^q$. The variable $z$ is to be interpreted as a $U(1)$ gauge symmetry fugacity. We recognize the contribution $z^{k'_+q}x^{(1-r)\frac{|q|}{2}}$ as the factors of the bare BPS monopole, which has gauge charge $G=k'_+ q$, dimension $\Delta =\frac{1-r}{2}|q|$ and angular momentum  $J_3 =0$. The factor $(1 - z x^{|q| + r + 2n})^{-1}$ matches the contribution of the BPS modes $\phi_{jj}$, with $j=\frac{|q|}{2}+n$. Indeed these modes obey $\Delta + J_3 = 2j +r = |q| + r + 2n$, and have gauge charge $G =1$. Similarly the factor $(1- z^{-1} x^{|q|+2-r + 2n})$ matches the contribution of the fermionic BPS mode $\ol\psi_{jj}$, with $j=\frac{|q|}{2}+\frac{1}{2}+n$, which have $\Delta + J_3 = 2j +1-r= |q| +2-r + 2n$ and gauge charge $G=-1$.
In the sector of magnetic charge $q$, the index receives contribution from gauge invariant BPS monopoles, which are associated with the terms of order $z^{-1}$ in the Laurent expansion in powers of $z$ of the integrand. It is not hard to see that this reproduces the constraint from Gauss's law $k'_+q = \sum_{j\ge \frac{|q|}{2}} n_{jj} - \sum_{j\ge \frac{|q|}{2}+\frac 12} \wat n_{jj}$, where $n_{jj} \in \bZ_{\ge 0}$ counts the dressing by bosonic modes $\phi_{jj}$ and $\wat n_{jj} \in \{0,1\}$ counts the dressing by fermionic modes $\ol\psi_{jj}$. 

We conclude that the BPS monopoles that we have described cover all the BPS local operators appearing in the superconformal index.\footnote{In the sector of zero magnetic charge $q=0$, the index simply counts the chiral operators which are gauge invariant polynomials of $\phi_{jj} \sim \p^j\phi$ and $\ol\psi_{jj} \sim \p^j \ol\psi$ (meaning the component with $J_3=j$).}
\footnote{In principle we expect other BPS monopoles whose contribution to the index cancel. These would be constructed with gaugino dressings and derivatives of the bare BPS monopole operator (whose precise meaning needs to be defined). It would be interesting to clarify this point in a future work.}

\subsection{Extensions and ABJM monopoles}
\label{ssec:Extensions}

It is straightforward to extend the discussion to SQED theories with any number $N_f$ of chiral multiplets, with various gauge charges. The construction of monopoles can a priori be extended to non-abelian theories without conceptual novelty, however the match with the superconformal index is less trivial in this case, as there can be cancelations between the contributions of different monopoles. As observed in \cite{Aharony:2015pla}, in non-abelian $\cN=2$ Chern-Simons theories there are monopoles dressed with gaugino modes which contribute oppositely to the index as the same monopoles without the gaugino dressing. More precisely, monopoles can recombine into long multiplets, which do not contribute to the index. Indeed the $A_1$ (or $\ol A_1$) superconformal multiplets to which the BPS monopoles belong are short multiplets {\it at threshold}, but are not {\it isolated}, in the language of \cite{Cordova:2016emh}, and two $A_1$ multiplets with appropriate quantum numbers can recombine.

\bigskip

Finally we would like make a comment on BPS monopoles in quiver Chern-Simons theories such as the ABJM theory \cite{Aharony:2008ug}. In the ABJM theory the gauge group is $U(N)_k\times U(N)_{-k}$, with the subscript indicating the Chern-Simons levels in each node, and the matter fields are four bifundamental chiral multiplets: $A_1,A_2$ in the representation $(N,\ol N)$ and $B_1,B_2$ in the representation $(\ol N,N)$. The theory has $\cN=6$ supersymmetry but we can regard it as an $\cN=2$ theory. Let us think about the abelian theory, with gauge group $U(1)_k\times U(1)_{-k}$ for simplicity. A bare monopole $v_{q_1,q_2}$ has gauge charge $(kq_1,-kq_2)$ under $U(1)_k\times U(1)_{-k}$. To built a gauge invariant operator one must dress the monopole with matter modes. Because matter fields are in (anti)bifundamental representations, this can be done only for bare monopoles of the form $v_{q,q}$. In that case a bifundamental field sees an effectively vanishing magnetic charge\footnote{In general a field with charges $(1,-1)$ under  $U(1)^2$ in the background of a $(q_1,q_2)$ monopole couples only to a $U(1)$ monopole of magnetic charge $q_1-q_2$.}, it is not charged under the $U(1) \subset U(N)^2$ that has the monopole singularity. Therefore the monopole singularity is dressed with standard insertions $\p^j A_\alpha(x),\p^j B_{\alpha}(x)$. One finds half-BPS (chiral) monopole operators $V^{+}_{\alpha_1\cdots \alpha_{kq}} = v_{q,q} B_{\alpha_1}\cdots B_{\alpha_{kq}}$ for $kq>0$, and $V^{-}_{\alpha_1\cdots \alpha_{kq}} = v_{q,q} A_{\alpha_1}\cdots A_{\alpha_{|kq|}}$ for $kq <0$, with zero spin.
This was described in \cite{Cremonesi:2016nbo}. 

We find that, despite having Chern-Simons terms, the monopoles in ABJM theory are half-BPS chiral operators. In particular they have no spin. This phenomenon carries on to other quiver Chern-Simons theories of a similar nature and in particular in Chern-Simons theories with extended $\cN \ge 3$ supersymmetry, as studied in \cite{Assel:2017eun}. In these theories the Chern-Simons levels and matter content are constrained in such a way as to allow for such half-BPS monopoles \cite{Gaiotto:2008sd,Hosomichi:2008jd} (in addition to $\frac 14$-BPS monopoles).


\appendix

\section{Supersymmetry transformations}
\label{app:susytransfo}

In this appendix we review the 3d $\cN=2$  supersymmetry transformations. We extract them from the $S^3$ supersymmetry tranformations of \cite{Hama:2010av} by taking the flat space limit.
For the abelian vector multiplet we have
\bea
&\delta A_\mu = -\frac{i}{2} ( \ol\varepsilon \gamma_\mu\lambda - \ol\lambda\gamma_\mu\varepsilon) \cr
&\delta\sigma = \frac 12 (\ol\varepsilon\lambda - \ol\lambda\varepsilon) \cr
&\delta\lambda = \frac{i}{2} \epsilon^{\mu\nu\rho} \gamma_\mu\varepsilon F_{\nu\rho} - D \varepsilon + i \gamma^\mu\varepsilon \p_\mu\sigma \cr
&\delta\ol\lambda = \frac{i}{2} \epsilon^{\mu\nu\rho} \gamma_\mu\ol\varepsilon F_{\nu\rho} + D \ol\varepsilon - i \gamma^\mu\ol\varepsilon \p_\mu\sigma \cr
& \delta D = -\frac i2 \ol\varepsilon\gamma^\mu\nabla_\mu\lambda + \frac i2 \varepsilon\gamma^\mu\nabla_\mu\ol\lambda 
\eea
and for the (anti)chiral multiplet we have
\bea
& \delta\phi = \ol\varepsilon\psi \cr
& \delta\psi = i \gamma^\mu\varepsilon D_\mu\phi + i\varepsilon \sigma\phi + \ol\varepsilon F \cr
& \delta F = \varepsilon (i\gamma^\mu D_\mu\psi - i \sigma\psi - i \lambda\phi) \cr
& \delta\ol\phi = \varepsilon\ol\psi \cr
& \delta\ol\psi = i \gamma^\mu\ol\varepsilon D_\mu\ol\phi + i\ol\varepsilon \ol\phi \sigma + \varepsilon \ol F \cr
& \delta \ol F = \ol\varepsilon (i\gamma^\mu D_\mu\ol\psi - i \ol\psi \sigma + i \ol\phi \ol\lambda) \,.
\eea

\bibliography{Benbib}

\providecommand{\href}[2]{#2}\begingroup\raggedright\begin{thebibliography}{10}

\bibitem{Borokhov:2002ib}
V.~Borokhov, A.~Kapustin and X.-k. Wu, \emph{{Topological disorder operators in
  three-dimensional conformal field theory}},
  \href{http://dx.doi.org/10.1088/1126-6708/2002/11/049}{\emph{JHEP} {\bf 11}
  (2002) 049}, [\href{https://arxiv.org/abs/hep-th/0206054}{{\tt
  hep-th/0206054}}].

\bibitem{Borokhov:2002cg}
V.~Borokhov, A.~Kapustin and X.-k. Wu, \emph{{Monopole operators and mirror
  symmetry in three-dimensions}},
  \href{http://dx.doi.org/10.1088/1126-6708/2002/12/044}{\emph{JHEP} {\bf 12}
  (2002) 044}, [\href{https://arxiv.org/abs/hep-th/0207074}{{\tt
  hep-th/0207074}}].

\bibitem{Borokhov:2003yu}
V.~Borokhov, \emph{{Monopole operators in three-dimensional N=4 SYM and mirror
  symmetry}},
  \href{http://dx.doi.org/10.1088/1126-6708/2004/03/008}{\emph{JHEP} {\bf 03}
  (2004) 008}, [\href{https://arxiv.org/abs/hep-th/0310254}{{\tt
  hep-th/0310254}}].

\bibitem{Read:1989zz}
N.~Read and S.~Sachdev, \emph{{Valence-bond and spin-Peierls ground states of
  low-dimensional quantum antiferromagnets}},
  \href{http://dx.doi.org/10.1103/PhysRevLett.62.1694}{\emph{Phys. Rev. Lett.}
  {\bf 62} (1989) 1694--1697}.

\bibitem{Read:1990zza}
N.~Read and S.~Sachdev, \emph{{Spin-Peierls, valence-bond solid, and Neel
  ground states of low-dimensional quantum antiferromagnets}},
  \href{http://dx.doi.org/10.1103/PhysRevB.42.4568}{\emph{Phys. Rev.} {\bf B42}
  (1990) 4568--4589}.

\bibitem{Senthil1490}
T.~Senthil, A.~Vishwanath, L.~Balents, S.~Sachdev and M.~P.~A. Fisher,
  \emph{Deconfined quantum critical points},
  \href{http://dx.doi.org/10.1126/science.1091806}{\emph{Science} {\bf 303}
  (2004) 1490--1494}.

\bibitem{2004PhRvB..70n4407S}
T.~{Senthil}, L.~{Balents}, S.~{Sachdev}, A.~{Vishwanath} and M.~P.~A.
  {Fisher}, \emph{{Quantum criticality beyond the Landau-Ginzburg-Wilson
  paradigm}},  \href{https://arxiv.org/abs/cond-mat/0312617}{{\tt
  cond-mat/0312617}}.

\bibitem{Aharony:2015mjs}
O.~Aharony, \emph{{Baryons, monopoles and dualities in Chern-Simons-matter
  theories}}, \href{http://dx.doi.org/10.1007/JHEP02(2016)093}{\emph{JHEP} {\bf
  02} (2016) 093}, [\href{https://arxiv.org/abs/1512.00161}{{\tt 1512.00161}}].

\bibitem{Seiberg:2016gmd}
N.~Seiberg, T.~Senthil, C.~Wang and E.~Witten, \emph{{A Duality Web in 2+1
  Dimensions and Condensed Matter Physics}},
  \href{http://dx.doi.org/10.1016/j.aop.2016.08.007}{\emph{Annals Phys.} {\bf
  374} (2016) 395--433}, [\href{https://arxiv.org/abs/1606.01989}{{\tt
  1606.01989}}].

\bibitem{Benini:2017dus}
F.~Benini, P.-S. Hsin and N.~Seiberg, \emph{{Comments on global symmetries,
  anomalies, and duality in (2 + 1)d}},
  \href{http://dx.doi.org/10.1007/JHEP04(2017)135}{\emph{JHEP} {\bf 04} (2017)
  135}, [\href{https://arxiv.org/abs/1702.07035}{{\tt 1702.07035}}].

\bibitem{Komargodski:2017keh}
Z.~Komargodski and N.~Seiberg, \emph{{A symmetry breaking scenario for
  QCD$_{3}$}}, \href{http://dx.doi.org/10.1007/JHEP01(2018)109}{\emph{JHEP}
  {\bf 01} (2018) 109}, [\href{https://arxiv.org/abs/1706.08755}{{\tt
  1706.08755}}].

\bibitem{Benini:2017aed}
F.~Benini, \emph{{Three-dimensional dualities with bosons and fermions}},
  \href{http://dx.doi.org/10.1007/JHEP02(2018)068}{\emph{JHEP} {\bf 02} (2018)
  068}, [\href{https://arxiv.org/abs/1712.00020}{{\tt 1712.00020}}].

\bibitem{Jensen:2017bjo}
K.~Jensen, \emph{{A master bosonization duality}},
  \href{http://dx.doi.org/10.1007/JHEP01(2018)031}{\emph{JHEP} {\bf 01} (2018)
  031}, [\href{https://arxiv.org/abs/1712.04933}{{\tt 1712.04933}}].

\bibitem{Benini:2018umh}
F.~Benini and S.~Benvenuti, \emph{{$\mathcal{N}{=}1$ dualities in 2+1
  dimensions}},  \href{https://arxiv.org/abs/1803.01784}{{\tt 1803.01784}}.

\bibitem{Cremonesi:2013lqa}
S.~Cremonesi, A.~Hanany and A.~Zaffaroni, \emph{{Monopole operators and Hilbert
  series of Coulomb branches of $3d$ $\mathcal{N} = 4$ gauge theories}},
  \href{http://dx.doi.org/10.1007/JHEP01(2014)005}{\emph{JHEP} {\bf 01} (2014)
  005}, [\href{https://arxiv.org/abs/1309.2657}{{\tt 1309.2657}}].

\bibitem{Affleck:1982as}
I.~Affleck, J.~A. Harvey and E.~Witten, \emph{{Instantons and (Super)Symmetry
  Breaking in (2+1)-Dimensions}},
  \href{http://dx.doi.org/10.1016/0550-3213(82)90277-2}{\emph{Nucl. Phys.} {\bf
  B206} (1982) 413--439}.

\bibitem{Aharony:1997bx}
O.~Aharony, A.~Hanany, K.~A. Intriligator, N.~Seiberg and M.~Strassler,
  \emph{{Aspects of N=2 supersymmetric gauge theories in three-dimensions}},
  \href{http://dx.doi.org/10.1016/S0550-3213(97)00323-4}{\emph{Nucl.Phys.} {\bf
  B499} (1997) 67--99}, [\href{https://arxiv.org/abs/hep-th/9703110}{{\tt
  hep-th/9703110}}].

\bibitem{Aharony:1997gp}
O.~Aharony, \emph{{IR duality in d = 3 N=2 supersymmetric USp(2N(c)) and
  U(N(c)) gauge theories}},
  \href{http://dx.doi.org/10.1016/S0370-2693(97)00530-3}{\emph{Phys. Lett.}
  {\bf B404} (1997) 71--76}, [\href{https://arxiv.org/abs/hep-th/9703215}{{\tt
  hep-th/9703215}}].

\bibitem{Benvenuti:2016wet}
S.~Benvenuti and S.~Pasquetti, \emph{{3d $ \mathcal{N} $ = 2 mirror symmetry,
  pq-webs and monopole superpotentials}},
  \href{http://dx.doi.org/10.1007/JHEP08(2016)136}{\emph{JHEP} {\bf 08} (2016)
  136}, [\href{https://arxiv.org/abs/1605.02675}{{\tt 1605.02675}}].

\bibitem{Benini:2017dud}
F.~Benini, S.~Benvenuti and S.~Pasquetti, \emph{{SUSY monopole potentials in
  2+1 dimensions}},
  \href{http://dx.doi.org/10.1007/JHEP08(2017)086}{\emph{JHEP} {\bf 08} (2017)
  086}, [\href{https://arxiv.org/abs/1703.08460}{{\tt 1703.08460}}].

\bibitem{Benvenuti:2017kud}
S.~Benvenuti and S.~Giacomelli, \emph{{Abelianization and sequential
  confinement in $2+1$ dimensions}},
  \href{http://dx.doi.org/10.1007/JHEP10(2017)173}{\emph{JHEP} {\bf 10} (2017)
  173}, [\href{https://arxiv.org/abs/1706.04949}{{\tt 1706.04949}}].

\bibitem{Benini:2018bhk}
F.~Benini and S.~Benvenuti, \emph{{$N=1$ QED in 2+1 dimensions: Dualities and
  enhanced symmetries}},  \href{https://arxiv.org/abs/1804.05707}{{\tt
  1804.05707}}.

\bibitem{Chester:2017vdh}
S.~M. Chester, L.~V. Iliesiu, M.~Mezei and S.~S. Pufu, \emph{{Monopole
  Operators in $U(1)$ Chern-Simons-Matter Theories}},
  \href{http://dx.doi.org/10.1007/JHEP05(2018)157}{\emph{JHEP} {\bf 05} (2018)
  157}, [\href{https://arxiv.org/abs/1710.00654}{{\tt 1710.00654}}].

\bibitem{Wu:1976ge}
T.~T. Wu and C.~N. Yang, \emph{{Dirac Monopole Without Strings: Monopole
  Harmonics}},
  \href{http://dx.doi.org/10.1016/0550-3213(76)90143-7}{\emph{Nucl. Phys.} {\bf
  B107} (1976) 365}.

\bibitem{Wu:1977qk}
T.~T. Wu and C.~N. Yang, \emph{{Some Properties of Monopole Harmonics}},
  \href{http://dx.doi.org/10.1103/PhysRevD.16.1018}{\emph{Phys. Rev.} {\bf D16}
  (1977) 1018--1021}.

\bibitem{Cordova:2016emh}
C.~Cordova, T.~T. Dumitrescu and K.~Intriligator, \emph{{Multiplets of
  Superconformal Symmetry in Diverse Dimensions}},
  \href{https://arxiv.org/abs/1612.00809}{{\tt 1612.00809}}.

\bibitem{Moore:1989yh}
G.~W. Moore and N.~Seiberg, \emph{{Taming the Conformal Zoo}},
  \href{http://dx.doi.org/10.1016/0370-2693(89)90897-6}{\emph{Phys. Lett.} {\bf
  B220} (1989) 422--430}.

\bibitem{Pufu:2013vpa}
S.~S. Pufu, \emph{{Anomalous dimensions of monopole operators in
  three-dimensional quantum electrodynamics}},
  \href{http://dx.doi.org/10.1103/PhysRevD.89.065016}{\emph{Phys. Rev.} {\bf
  D89} (2014) 065016}, [\href{https://arxiv.org/abs/1303.6125}{{\tt
  1303.6125}}].

\bibitem{Chester:2015wao}
S.~M. Chester, M.~Mezei, S.~S. Pufu and I.~Yaakov, \emph{{Monopole operators
  from the $4-\epsilon$ expansion}},
  \href{http://dx.doi.org/10.1007/JHEP12(2016)015}{\emph{JHEP} {\bf 12} (2016)
  015}, [\href{https://arxiv.org/abs/1511.07108}{{\tt 1511.07108}}].

\bibitem{Radicevic:2015yla}
D.~Radicevic, \emph{{Disorder Operators in Chern-Simons-Fermion Theories}},
  \href{http://dx.doi.org/10.1007/JHEP03(2016)131}{\emph{JHEP} {\bf 03} (2016)
  131}, [\href{https://arxiv.org/abs/1511.01902}{{\tt 1511.01902}}].

\bibitem{Dyer:2013fja}
E.~Dyer, M.~Mezei and S.~S. Pufu, \emph{{Monopole Taxonomy in Three-Dimensional
  Conformal Field Theories}},  \href{https://arxiv.org/abs/1309.1160}{{\tt
  1309.1160}}.

\bibitem{Atiyah:1975jf}
M.~F. Atiyah, V.~K. Patodi and I.~M. Singer, \emph{{Spectral asymmetry and
  Riemannian Geometry 1}},
  \href{http://dx.doi.org/10.1017/S0305004100049410}{\emph{Math. Proc.
  Cambridge Phil. Soc.} {\bf 77} (1975) 43}.

\bibitem{Atiyah:1976jg}
M.~F. Atiyah, V.~K. Patodi and I.~M. Singer, \emph{{Spectral asymmetry and
  Riemannian geometry 2}},
  \href{http://dx.doi.org/10.1017/S0305004100051872}{\emph{Math. Proc.
  Cambridge Phil. Soc.} {\bf 78} (1976) 405}.

\bibitem{Atiyah:1980jh}
M.~F. Atiyah, V.~K. Patodi and I.~M. Singer, \emph{{Spectral asymmetry and
  Riemannian geometry. III}},
  \href{http://dx.doi.org/10.1017/S0305004100052105}{\emph{Math. Proc.
  Cambridge Phil. Soc.} {\bf 79} (1976) 71--99}.

\bibitem{Witten:2015aba}
E.~Witten, \emph{{Fermion Path Integrals And Topological Phases}},
  \href{http://dx.doi.org/10.1103/RevModPhys.88.035001,
  10.1103/RevModPhys.88.35001}{\emph{Rev. Mod. Phys.} {\bf 88} (2016) 035001},
  [\href{https://arxiv.org/abs/1508.04715}{{\tt 1508.04715}}].

\bibitem{Imamura:2011su}
Y.~Imamura and S.~Yokoyama, \emph{{Index for three dimensional superconformal
  field theories with general R-charge assignments}},
  \href{http://dx.doi.org/10.1007/JHEP04(2011)007}{\emph{JHEP} {\bf 04} (2011)
  007}, [\href{https://arxiv.org/abs/1101.0557}{{\tt 1101.0557}}].

\bibitem{Benini:2011cma}
F.~Benini, C.~Closset and S.~Cremonesi, \emph{{Quantum moduli space of
  Chern-Simons quivers, wrapped D6-branes and AdS4/CFT3}},
  \href{http://dx.doi.org/10.1007/JHEP09(2011)005}{\emph{JHEP} {\bf 09} (2011)
  005}, [\href{https://arxiv.org/abs/1105.2299}{{\tt 1105.2299}}].

\bibitem{Jafferis:2010un}
D.~L. Jafferis, \emph{{The Exact Superconformal R-Symmetry Extremizes Z}},
  \href{http://dx.doi.org/10.1007/JHEP05(2012)159}{\emph{JHEP} {\bf 05} (2012)
  159}, [\href{https://arxiv.org/abs/1012.3210}{{\tt 1012.3210}}].

\bibitem{Minwalla:2011ma}
S.~Minwalla, P.~Narayan, T.~Sharma, V.~Umesh and X.~Yin, \emph{{Supersymmetric
  States in Large N Chern-Simons-Matter Theories}},
  \href{http://dx.doi.org/10.1007/JHEP02(2012)022}{\emph{JHEP} {\bf 02} (2012)
  022}, [\href{https://arxiv.org/abs/1104.0680}{{\tt 1104.0680}}].

\bibitem{Bhattacharya:2008zy}
J.~Bhattacharya, S.~Bhattacharyya, S.~Minwalla and S.~Raju, \emph{{Indices for
  Superconformal Field Theories in 3,5 and 6 Dimensions}},
  \href{http://dx.doi.org/10.1088/1126-6708/2008/02/064}{\emph{JHEP} {\bf 02}
  (2008) 064}, [\href{https://arxiv.org/abs/0801.1435}{{\tt 0801.1435}}].

\bibitem{Kapustin:2011jm}
A.~Kapustin and B.~Willett, \emph{{Generalized Superconformal Index for Three
  Dimensional Field Theories}},  \href{https://arxiv.org/abs/1106.2484}{{\tt
  1106.2484}}.

\bibitem{Aharony:2015pla}
O.~Aharony, P.~Narayan and T.~Sharma, \emph{{On monopole operators in
  supersymmetric Chern-Simons-matter theories}},
  \href{http://dx.doi.org/10.1007/JHEP05(2015)117}{\emph{JHEP} {\bf 05} (2015)
  117}, [\href{https://arxiv.org/abs/1502.00945}{{\tt 1502.00945}}].

\bibitem{Aharony:2008ug}
O.~Aharony, O.~Bergman, D.~L. Jafferis and J.~Maldacena, \emph{{N=6
  superconformal Chern-Simons-matter theories, M2-branes and their gravity
  duals}}, \href{http://dx.doi.org/10.1088/1126-6708/2008/10/091}{\emph{JHEP}
  {\bf 0810} (2008) 091}, [\href{https://arxiv.org/abs/0806.1218}{{\tt
  0806.1218}}].

\bibitem{Cremonesi:2016nbo}
S.~Cremonesi, N.~Mekareeya and A.~Zaffaroni, \emph{{The moduli spaces of 3d $
  \mathcal{N}\ge 2 $ Chern-Simons gauge theories and their Hilbert series}},
  \href{http://dx.doi.org/10.1007/JHEP10(2016)046}{\emph{JHEP} {\bf 10} (2016)
  046}, [\href{https://arxiv.org/abs/1607.05728}{{\tt 1607.05728}}].

\bibitem{Assel:2017eun}
B.~Assel, \emph{{The Space of Vacua of 3d $\mathcal{N}=3$ Abelian Theories}},
  \href{http://dx.doi.org/10.1007/JHEP08(2017)011}{\emph{JHEP} {\bf 08} (2017)
  011}, [\href{https://arxiv.org/abs/1706.00793}{{\tt 1706.00793}}].

\bibitem{Gaiotto:2008sd}
D.~Gaiotto and E.~Witten, \emph{{Janus Configurations, Chern-Simons Couplings,
  And The theta-Angle in N=4 Super Yang-Mills Theory}},
  \href{http://dx.doi.org/10.1007/JHEP06(2010)097}{\emph{JHEP} {\bf 1006}
  (2010) 097}, [\href{https://arxiv.org/abs/0804.2907}{{\tt 0804.2907}}].

\bibitem{Hosomichi:2008jd}
K.~Hosomichi, K.-M. Lee, S.~Lee, S.~Lee and J.~Park, \emph{{N=4 Superconformal
  Chern-Simons Theories with Hyper and Twisted Hyper Multiplets}},
  \href{http://dx.doi.org/10.1088/1126-6708/2008/07/091}{\emph{JHEP} {\bf 0807}
  (2008) 091}, [\href{https://arxiv.org/abs/0805.3662}{{\tt 0805.3662}}].

\bibitem{Hama:2010av}
N.~Hama, K.~Hosomichi and S.~Lee, \emph{{Notes on SUSY Gauge Theories on
  Three-Sphere}}, \href{http://dx.doi.org/10.1007/JHEP03(2011)127}{\emph{JHEP}
  {\bf 1103} (2011) 127}, [\href{https://arxiv.org/abs/1012.3512}{{\tt
  1012.3512}}].

\end{thebibliography}\endgroup
\bibliographystyle{JHEP}

\end{document}